\documentclass[preprint,aps,superscriptaddress,nofootinbib,showkeys,10pt]{revtex4}
\pdfoutput=1
\usepackage{graphicx}
\usepackage{amsmath}
\usepackage{amsfonts}
\usepackage{amssymb}
\usepackage{xcolor, soul}
\usepackage{epstopdf}
\usepackage{hyperref}
\usepackage{natbib}
\hypersetup{
  colorlinks   = true,
  citecolor    = red,
  linkcolor    = blue,
  urlcolor     = blue,
}
\def\bse{\begin{subequations}}
\def\ese{\end{subequations}}

\def\mpl{M_{pl}}
\def\wre{w_{re}}
\def\xre{x_{re}}

\def\HI{H_I}
\def\l{\left}
\def\r{\right}
\def\f {\frac}
\def\ae{a_{\rm e}}
\def\are{a_{\rm re}}
\def\ee{\eta_{\rm e}}

\def\xe{x_{\rm e}}
\def\mJ{\mathrm{J}}
\def\xre{x_{\rm re}}
\def\ere{\eta_{\rm re}}
\def\kre{k_{\rm re}}
\def\ke{k_{\rm e}}

\def\mpl{M_{\rm pl}}

\def\pb{\mathcal{P}_{\rm B}}
\def\pe{\mathcal{P}_{\rm E}}
\def\Pt{\mathcal{P}_{\rm T}}
\def\mPt{\mathcal{P}_{\rm T}}
\def\mPtp{\mathcal{P}_{\rm T}^{\rm pri}}
\def\mPts{\mathcal{P}_{\rm T}^{\rm sec}}
\def\mPtsra{\mathcal{P}_{\rm T, ra}^{\rm sec}}

\def\hk{{h}^{\lambda}_{\bf k}}

\def\xe{x_{\rm end}}
\def\l{\left}
\def\r{\right}

\def\Mp{M_{\rm P}}

\def\Gev{\mbox{ GeV}}

\def\nn{\nonumber}

\def\are{a_{\rm re}}

\def\Gk{\mathcal{G}_{k}}

\def\fre{f_{\rm re}}

\def\vq{\mathbf{q}}
\def\As{A_{\rm s}}

\def\kpv{k_{*}}

\def\ogwh{\Omega_{\rm gw}h^2}
\def\ogw{\Omega_{\rm gw}}
\def\ogw{\Omega_{_{\rm GW}}}
\def\ogwp{\Omega_{_{\rm GW}}^{\rm PRI}}

\def\f{\frac}

\def\mB{\mathcal{B}}
\def\ee{{\eta}_{\mathrm{end}}}

\def\vk{ \mathbf{k}}
\def\ere{\eta_{\rm re}}
\def\xre{x_{\rm re}}

\def\mIra{\mathcal{I}_{\rm}}

\def\knu{k_\nu}
\def\kre{k_{\rm re}}

\def\gre{g_{\rm re}}
\def\Tre{T_{\rm re}}

\def\Sem{S_{\rm EM}}
\def\vx{\mathbf{x}}
\def\vk{\mathbf{k}}
\def\vq{\mathbf{q}}
\def\Akl{A_{\vk}^{\lambda}}
\def\mPb{\mathcal{P}_{\rm B}}
\def\mPe{\mathcal{P}_{\rm E}}
\def\NI{N_{\rm I}}
\def\Nre{N_{\rm re}}
\def\mAk{\mathcal{A}_{\vk}}
\def\mAki{\mathcal{A}_{\vk}^{\mathrm{inf}}}
\def\Ak{A_{\vk}}
\def\bJ{\mathrm{J}}
\def\bY{\mathrm{Y}}
\def\rmH{\mathrm{H}}
\def\Mpc{\mathrm{Mpc}}
\def\kcmb{k_{*}}

\def\mGk{\mathcal{G}_k}
\def\mPbe{\mathcal{P}_{\rm B/E}}
\def\hkr{h_{k,\rm ra}^{\lambda}}
\def\hkrp{h_{k,\rm ra}^{\lambda '}}
\def\mI{\mathcal{I}}
\def\umin{u_{\rm min}}
\def\umax{u_{\rm max}}
\def\nw{n_w}
\def\mD{\mathcal{D}}
\def\gsp{g_{\rm s, 0}}
\def\gseq{g_{\rm s,eq}}

\graphicspath{{./figs/}}
\begin{document}

\title{Magnetogenesis from Sawtooth Coupling: Gravitational Wave Probe of Reheating \\
}
\author{Subhasis Maiti}
\email{E-mail: subhashish@iitg.ac.in}
\affiliation{Department of Physics, Indian Institute of Technology, Guwahati, 
Assam, India}

\date{\today}

\begin{abstract}

The detection of gravitational waves (GWs) by LIGO-Virgo and pulsar timing arrays (PTAs) has opened a new window into early universe cosmology. Yet, the origin of large-scale magnetic fields and the dynamics of the reheating epoch remain poorly understood. In this work, we study the generation of secondary GWs (SGWs) sourced by primordial magnetic fields produced via a Sawtooth-type coupling during reheating with a general background evolution.
We show that the reheating equation of state significantly influences the spectral shape and amplitude of the magnetic fields. While a scale-invariant spectrum is typically needed to match observational bounds, this coupling naturally produces a strongly blue-tilted spectrum that remains consistent with current constraints. Crucially, the magnetic field continues to grow during reheating, leading to a GW signal with a broken power-law spectrum and a distinctive blue tilt on super-horizon scales.
This SGW signal can fall within the sensitivity of upcoming detectors such as LISA, DECIGO, and BBO. The unique spectral features make this scenario distinguishable from other sources, offering a viable mechanism for cosmic magnetogenesis and a novel probe of the reheating era through GW observations.

\end{abstract}
\maketitle

\section{Introduction}

One of the most intriguing mysteries of the universe is its widespread magnetization from the very early stages of cosmic evolution. Observational evidence, both direct and indirect, supports the existence of magnetic fields over a vast range of strengths ($10^{-16}$ to $10^{15}$ G) and corresponding length scales \cite{Giovannini:2003yn, Grasso:2000wj, Kronberg:2001st, Durrer:2013pga, Subramanian:2015lua}. For instance, the Milky Way hosts a magnetic field of approximately $10^{-6}$ G, while galaxies and galaxy clusters exhibit field strengths on the order of a few $\mu$G \cite{Turner:1987bw, Grasso:2000wj, Giovannini:2003yn, Kronberg:2001st}. Furthermore, indirect measurements suggest that the intergalactic medium (IGM) and cosmic voids contain magnetic fields of about $10^{-16}$ G \cite{Neronov:2010gir, Tavecchio:2010mk, Dolag:2010ni, Dermer:2010mm,
Vovk:2011aa,Taylor:2011bn,Takahashi:2011ac}. In addition, the anisotropies in the cosmic microwave background (CMB) provide an 
upper bound on the strength of the primordial magnetic fields~(PMFs) to be of the 
order of $\mathrm{nG}$ on Mpc scales~\cite{paoletti2022constraints, Zucca:2016iur}. While astrophysical processes can account for the generation of magnetic fields in galaxies and clusters, the origin of magnetic fields in the IGM and voids remains an open question.  

Among the proposed mechanisms for generating such large-scale fields, a prominent cosmological scenario was introduced by Turner and Widrow, who suggested that inflation could serve as a natural magnetogenesis mechanism if the conformal invariance of the electromagnetic (EM) field is broken \cite{PhysRevD.37.2743}. Since standard Maxwell theory is conformally invariant, breaking this invariance can be achieved in different ways, coupling photons to the gravitational field $(R^n/\Mp^{2n}) FF$ \cite{PhysRevD.37.2743, Papanikolaou:2024cwr}, Kinatic coupling like $f^2(\phi)FF$ \cite{Demozzi:2009fu,Ratra:1991bn, Kandus:2010nw, PhysRevD.37.2743, Durrer:2013pga, Ferreira:2013sqa, Subramanian:2015lua, Kobayashi:2014sga, Haque:2020bip, Tripathy:2021sfb, Li:2022yqb, Maiti:2025cbi, Sharma:2017eps, PhysRevD.111.083550, Cecchini:2023bqu} or $f^2(\phi)F\tilde{F}$\cite{Campanelli:2008kh,Jain:2012jy, Caprini:2014mja, Sharma:2018kgs, Bamba:2021wyx, Sharma:2018kgs}, anomalous coupling of photons to axions $\phi\, F \tilde{F}$ \cite{Adshead:2016iae}.
Another potential mechanism for the generation of primordial magnetic fields arises from first-order phase transitions in the early universe~\cite{PhysRevD.53.662, 1998PhRvD..57..664A, Yang:2021uid}, such as the electroweak and quantum chromodynamics (QCD) phase transitions~\cite{Quashnock:1988vs, Fu:2010rs, PhysRevD.104.063508}. However, these mechanisms typically struggle to explain magnetic fields on large cosmological scales, as the relevant modes remain super-horizon during these transitions. 
If the phase transitions produce helical magnetic fields in the presence of turbulence, it becomes possible for magnetic energy to be transferred from smaller to larger scales via an inverse cascade process~\cite{PhysRevE.64.056405, Durrer:2013pga, Caprini:2009yp, Brandenburg:1996fc, Caprini:2001nb}.

Inflation provides a compelling mechanism for generating large-scale magnetic fields from quantum fluctuations \cite{Guth:1980zm, Linde:1981mu, Albrecht:1982wi, Starobinsky:1980te, PhysRevD.50.7222}. However, due to the conformal invariance of the standard electromagnetic (EM) action, magnetic fields produced during inflation typically decay rapidly as $B \propto 1/a^2$ with the universe’s expansion. To sustain significant field strengths, explicit breaking of conformal invariance is necessary. Several mechanisms have been proposed in the literature, but many suffer from severe backreaction effects or strong coupling issues \cite{Ferreira:2013sqa, Kobayashi:2014sga, PhysRevD.100.023524, Haque:2020bip, Maity:2021qps, Tripathy:2021sfb, Li:2022yqb, Benevides:2018mwx, Kobayashi:2019uqs, PhysRevD.94.043523, Hortua:2014wna, Campanelli:2008kh, Jain:2012jy, Caprini:2014mja, Sharma:2018kgs, Bamba:2021wyx, PhysRevD.37.2743, PhysRevD.94.043523}.
A key requirement for successful inflationary magnetogenesis is the production of a nearly scale-invariant magnetic spectrum at 1 Mpc \cite{Tripathy:2021sfb, Caprini:2014mja}. Although some models generate scale-dependent magnetic fields at large scales, achieving sufficient field strength typically requires inflationary scenarios with a relatively low energy scale \cite{Sharma:2017eps, Papanikolaou:2024cwr}. Many existing studies consider reheating phases, but often fail to explain the observed large-scale magnetic fields adequately.
Crucially, inflationary and reheating parameters are not independent. In realistic scenarios, the dynamics of reheating are closely tied to those of inflation. Even in the simplest perturbative reheating models, the inflationary parameters—such as the energy scale of inflation \( \HI \) and the number of e-folds \( \NI \) can be expressed in terms of two reheating parameters: the effective equation of state \( \wre \) and the reheating temperature \( \Tre \).

In this work, we systematically incorporate reheating dynamics, focusing on perturbative reheating scenarios using a model-independent approach described in \cite{PhysRevLett.113.041302}. Our analysis reveals that reheating significantly influences the evolution of the gauge field. We find that in most inflationary magnetogenesis scenarios, a reheating phase characterized by matter-like evolution and a low reheating temperature results in an extremely weak present-day magnetic field, failing to meet observational constraints. To resolve this, an additional mechanism is needed to enhance the large-scale magnetic field after inflation \cite{Kobayashi:2019uqs, Haque:2020bip, Maiti:2025cbi}. Interestingly, we demonstrate that even though the generated magnetic field exhibits strong scale dependence, a prolonged reheating phase with a sufficiently low temperature can still yield magnetic field strengths consistent with observations.

We explore a scenario where the coupling responsible for magnetogenesis remains active during reheating. While previous studies have investigated similar models, they often assume a very low inflationary energy scale \cite{ Sharma:2017eps, Papanikolaou:2024cwr} and neglect detailed reheating dynamics. Given that the reheating phase is closely tied to the post-inflationary behavior of the inflaton field \cite{PhysRevD.82.023511, PhysRevD.98.103525, PhysRevD.107.043531}, it is essential to incorporate a realistic reheating scenario when analyzing magnetogenesis. Our goal is to provide a comprehensive framework that connects inflationary magnetogenesis with reheating physics, ensuring a consistent treatment of both phases.

Although large-scale magnetic fields are primarily inferred through indirect observational evidence, their presence in the early universe could lead to the production of secondary gravitational waves (GWs). With state-of-the-art observatories designed to detect the stochastic gravitational wave background (SGWB), we now have an unprecedented opportunity to probe inflationary magnetogenesis directly. Since GWs interact weakly with the cosmic background, they serve as pristine messengers, carrying crucial information about inflation and the subsequent evolution of the universe.

We analyze how inflationary electromagnetic fields can generate a significant SGWB signal, potentially detectable by ongoing GW experiments such as pulsar timing arrays (PTAs) \cite{Lentati:2015qwp, NANOGrav:2020bcs, NANOGrav:2023gor, 2023arXiv230616224A, Reardon:2023gzh, Zic:2023gta, Xu:2023wog} and the Square Kilometre Array (SKA) \cite{Janssen:2014dka}, as well as future GW observatories like LISA \cite{Amaro-Seoane:2012aqc, Barausse:2020rsu}, BBO \cite{Crowder:2005nr, Corbin:2005ny, Baker:2019pnp}, DECIGO \cite{Seto:2001qf, Kawamura:2011zz, Suemasa:2017ppd}, and the Einstein Telescope (ET) \cite{Punturo:2010zz, Sathyaprakash:2012jk} or in $\mu$Hz frequency rage detectors~\cite{Blas:2021mqw, Foster:2025nzf}. Our findings highlight the potential of GW observations as a powerful probe of inflationary magnetogenesis and the post-inflationary evolution of the universe. 

This paper is structured as follows: i)In Section~\ref{sec:magnetogenesis}, we introduce the magnetogenesis model based on the coupling $f^2(\phi) F_{\mu\nu} F^{\mu\nu}$, discussing how a modified coupling function can simultaneously mitigate both the strong coupling and backreaction problems. Additionally, we explore the role of perturbative reheating and analyze how the magnetic spectral properties depend on reheating dynamics.
    ii) In Section~\ref{sec:GWs}, we investigate the detectability of these models using GWs and demonstrate that such scenarios can generate strong GW signals in the nano-Hz frequency range, potentially observable by future GW experiments.
   iii)  Finally, in Section~\ref{sec:conclusion}, we summarize our key findings, emphasizing that this model successfully explains the origin of large-scale magnetic fields without encountering the usual theoretical challenges while also predicting significant GW signals within detectable frequency ranges.

\section{Evolution of the Gauge Fields}\label{sec:magnetogenesis}
The generation of primordial magnetic fields (PMFs) requires a mechanism to break the conformal invariance of Maxwell's equations in the expanding universe, thereby allowing the excitation of gauge fields from the quantum vacuum. Several approaches have been proposed to achieve this symmetry breaking. These include direct couplings of the gauge field to the inflationary sector or other scalar fields, such as interactions of the form \( f(\phi) F_{\mu\nu} \tilde{F}^{\mu\nu} \)~\cite{Campanelli:2008kh, Jain:2012jy, Caprini:2014mja, Sharma:2018kgs, Bamba:2021wyx}; non-minimal couplings with the Ricci scalar, such as \( \xi R A_\mu A^\mu \)~\cite{PhysRevD.37.2743}; or axion-like interactions of the form  $\frac{\alpha \chi}{f_a} F_{\mu\nu} \tilde{F}^{\mu\nu}$ ~\cite{Adshead:2016iae, Adshead:2016iae, PhysRevLett.121.031102, Patel:2019isj, Brandenburg:2024awd}.  

In this work, we focus on a specific type of coupling, namely the \( f(\phi) F F \) interaction, with certain modifications that simultaneously address the strong coupling problem and the backreaction issue (see, for instance,~\cite{Sharma:2017eps}). Importantly, this mechanism is capable of successfully explaining the large-scale magnetic fields observed in the cosmic microwave background (CMB)~\cite{paoletti2022constraints, Zucca:2016iur, BICEP2:2017lpa} and in Fermi/LAT and HESS observations of TeV blazars (see, for example,~\cite{Neronov:2010gir, Tavecchio:2010mk, Dolag:2010ni, Dermer:2010mm, Vovk:2011aa,
Taylor:2011bn,Takahashi:2011ac}).
Refs.). 

Now, to break the conformal symmetry, we introduced a time-dependent effective gauge coupling $f(\eta)$ is typically introduced \cite{Kobayashi:2014sga, Ferreira:2013sqa, Subramanian:2015lua, Haque:2020bip, Tripathy:2021sfb, Maiti:2025cbi}.
\begin{align}
\Sem=
-\frac{1}{4}\int d^4x\sqrt{-g}g^{\alpha\beta}g^{\mu\nu}f^2(\eta)F_{\mu\alpha}F_{\nu\beta} .\label{eq:action_1}
\end{align}
where \( F_{\mu\nu} = \partial_{\mu} A_{\nu} - \partial_{\nu} A_{\mu} \) is the electromagnetic field tensor, and \( A_{\mu} \) is the four-vector potential. For simplicity, we restrict our analysis to cases where the total produced magnetic field remains smaller than the background energy density, ensuring that our model remains free from any backreaction issues.  

For a spatially flat FLRW metric, the background in conformal coordinates is given by  
\begin{align}\label{metric-element}
ds^2 = g_{\mu\nu} dx^{\mu} dx^{\nu} = a^2(\eta) (-d\eta^2 + dx^2),
\end{align}  
where \( \eta \) is the conformal time. The assumption of spatial flatness allows us to express the vector potential in terms of irreducible components as  $A_\mu = \left(A_0, \partial_i S + A_i \right)$,   
with the traceless condition $\delta^{ij} \partial_i A_j = 0$.  
In terms of these components, the action in Eq.~\eqref{eq:action_1} simplifies to
\begin{equation}
\label{action_pot}
\Sem=\frac{1}{2}\int d\eta d^3x f(\eta)^2 (A_i^\prime {A^i}^\prime-\partial_i A_j\partial^i A^j).
\end{equation}
Due to the inherent conformal invariance, the action becomes independent of the scale factor. The spatial indices are raised or lowered using the usual Kronecker delta function. Assuming the Fourier expansion of \( A_i \) as  
\begin{equation}
\label{mode_EM}
A_i(\eta,\vx)=\sum_{\lambda=1,2}\int\frac{d^3k}{(2\pi)^3}\epsilon^{\lambda}_i(\vk)  e^{i\vk\cdot\vx}\Akl(\eta),
\end{equation}
with the reality condition \( A^{\lambda}_{-\vk} = A^{\lambda*}_{\vk} \), where the polarization vector \( \epsilon_i^{\lambda}(\vk) \) corresponds to the two modes \( \lambda = 1,2 \) and satisfies  $\epsilon_i^{\lambda}(\vk) k_i = 0, \quad \text{and} \quad \epsilon_i^{\lambda}(\vk) \epsilon_i^{\lambda'}(\vk) = \delta_{\lambda\lambda'}.$ 
In this scenario, a non-helical magnetic field is generally generated, where both modes are equally excited from the vacuum due to the coupling (for instance, see~\cite{Maiti:2025cbi}).  

Now, we can write the equation of motion (EoM) for the associated mode function of the gauge field \( \Akl \) as~\cite{Haque:2020bip, Tripathy:2021sfb},
\begin{align}\label{eq:Ak_lambda}
    {\Akl}''+ 2\frac{f^\prime}{f}{\Akl}'+k^2\Akl=0
\end{align}
Here, the overprime \( (') \) denotes the derivative with respect to conformal time \( \eta \). Now, we can define the spectral energy density in the magnetic and electric fields as~\cite{Subramanian:2015lua}
\begin{align}\label{eq:def_pbe}
    \mPb(k,\eta) &=\frac{f^2}{2\pi^2}\l(\frac{k}{a}\r)^4k\sum_{\lambda}\l|\Akl(\eta)\r|^2=\frac{k^5}{2\pi^2a^4(\eta)}\sum_{\lambda}\l|\mAk^{\lambda}(\eta)\r|^2\\
    \mPe(k,\eta) &=\frac{f^2}{2\pi^2}\frac{k^3}{a^4}\sum_{\lambda}\l|{\Akl}'(\eta)\r|^2=\frac{k^3}{2\pi^2a^4(\eta)}\sum_{\lambda} \l|{\mAk^{\lambda}}'(\eta)-\frac{f'}{f}\mAk^{\lambda}(\eta)\r|^2
\end{align}
where we define \( \mAk^\lambda = f \Akl \).  

To determine the spectral behavior of the magnetic field arising from this specific coupling, we first need to understand how the coupling function evolves during the early stages of the universe. In the following section, we discuss the nature of the coupling function \( f(\phi) \).

\paragraph{\underline{Functional behavior of the coupling function:}\\}  
Here, we consider a broken power-law behavior for the coupling function. Initially, during inflation, the coupling function increases with time following a power-law behavior, \( f(a) \propto a^n \). After inflation, it decreases with time following another power-law behavior, \( f(a) \propto a^{-m} \). The coupling function can be expressed as  
\begin{align}
    f(a)=\l\{
    \begin{matrix}
        \l(\frac{a}{a_i}\r)^n & a_i\leq a\leq\ae\\
       \l(\frac{\ae}{a_i}\r)^n \l(\frac{\ae}{a}\r)^m & \ae\leq a\leq\are\\
       1 & a>\are
    \end{matrix}
    \r.
\end{align}
Here, \( \ae \) and \( \are \) denote the scale factors at the end of inflation and reheating, respectively. In a de Sitter inflationary background, the evolution of the scale factor is given by  $a(\eta) = -{1}/{\HI \eta}$.
During reheating, the scale factor evolves according to the general equation of state (EoS, \( \wre \)) as  $a(\eta > \eta_e) = \ae \left( {\eta}/{\eta_e} \right)^{\delta}$,
where \( \delta(\wre) = 2/(1 + 3\wre) \).  

\begin{figure*}
\includegraphics[width=0.45\linewidth]{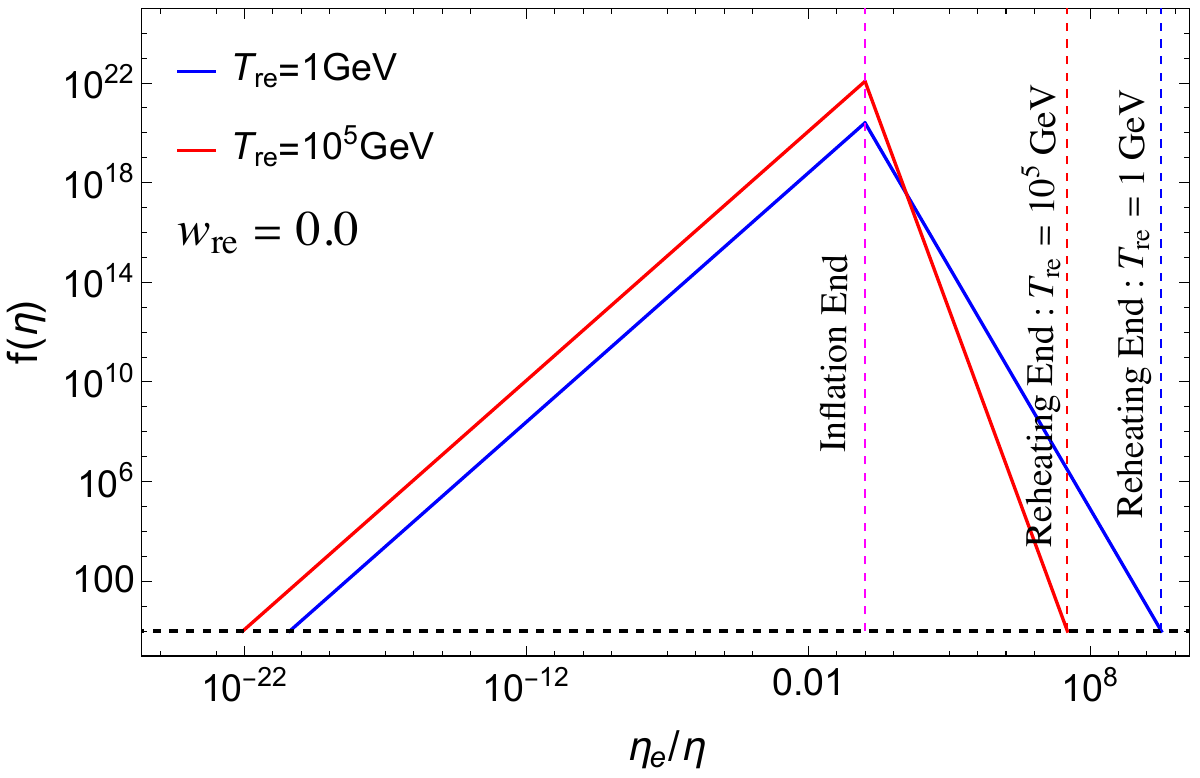}
\includegraphics[width=0.45\linewidth]{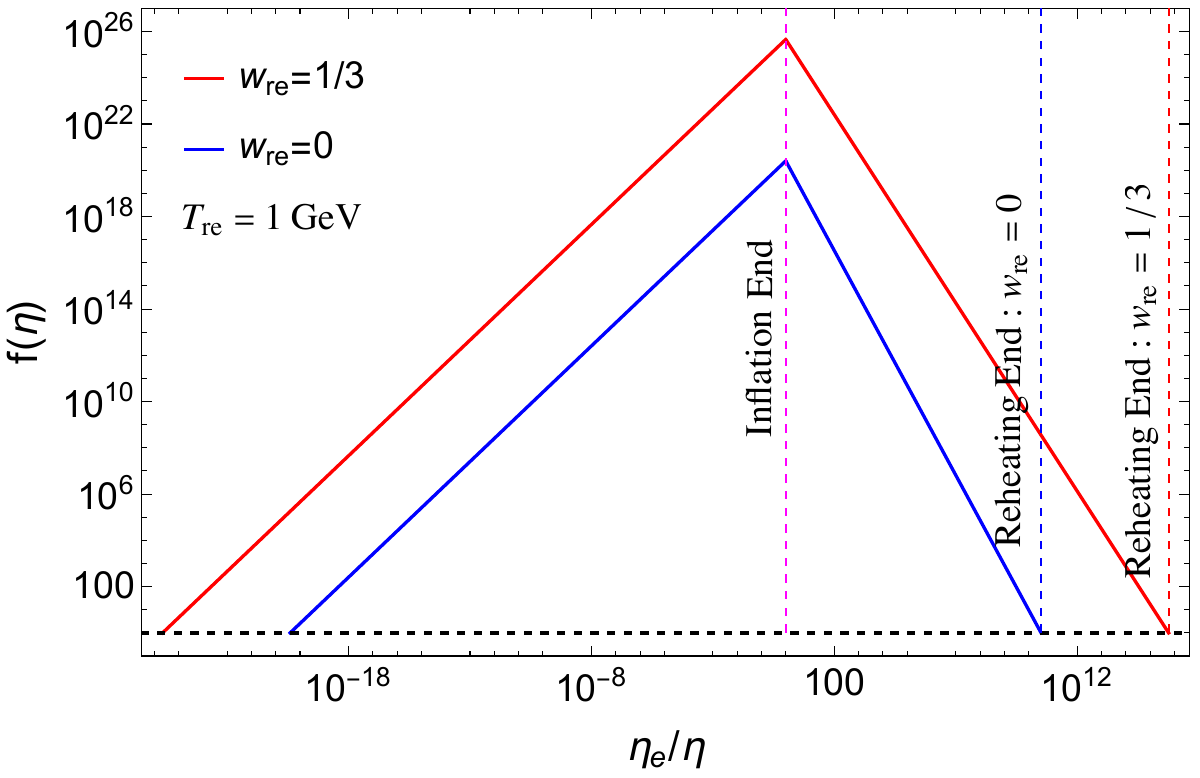}
\caption{In the above figure, we illustrate the evolution of the coupling function \( f(\eta) \) during the early stages of the universe, plotted as a function of \( \ee/\eta \). In the left panel, we assume a constant equation of state \( \wre = 0 \), and show two different reheating temperatures: \( \Tre = 1\,\Gev \) (blue) and \( \Tre = 10^5\,\Gev \) (red). In the right panel, we fix the reheating temperature at \( \Tre = 1\,\Gev \), and compare two different equations of state: \( \wre = 0 \) (blue) and \( \wre = 1/3 \) (red). In both panels, the black dashed horizontal line indicates the value \( f(\eta) = 1 \).}
    \label{fig:f_vs_eta}
\end{figure*}

Now, expressing the coupling function in terms of conformal time, we obtain  
\begin{align}
    f(\eta) \propto \l\{
    \begin{matrix}
        \left(\frac{\eta_i}{\eta}\right)^n & \quad a_i \leq a \leq \ae, \\
        \left(\frac{\eta_e}{\eta}\right)^\alpha & \quad \ae \leq a \leq \are,\\
        1 & a>\are
    \end{matrix}
    \r.
\end{align}  
where \( \alpha = {2n\beta}/{(1 + 3\wre)} \), and we define \( \beta = {\NI}/{\Nre} \). Here, \( \NI \) and \( \Nre \) represent the e-folding numbers of the inflationary and reheating eras, respectively. This relation must hold to restore the conformal nature of the gauge field at the end of reheating. In Fig.~\ref{fig:f_vs_eta}, we show the evolution of the coupling function \( f(\eta) \) during inflation and the subsequent reheating phase for two different scenarios. In the left panel, we fix the equation of state to \( \wre = 0 \), and plot the behavior of \( f(\eta) \) for two different reheating temperatures, \( \Tre = 1\,\Gev \) (blue) and \( \Tre = 10^5\,\Gev \) (red) as function of conformal time $\eta$. In the right panel, we fix the reheating temperature to \( \Tre = 1\,\Gev \), and illustrate the effect of two different equations of state, \( \wre = 0 \) (blue) and \( \wre = 1/3 \) (red). In both panels, we set \( n = 1.0 \). 
Now, we consider this functional form of the coupling to solve both the strong coupling and backreaction issues. We will discuss how this type of coupling is free from both of these.  
Let us first address the backreaction issue. If we work in the regime where \( 0 < n \leq 2 \), then both the electric and magnetic spectral energy densities, \( \mPb(k) \) and \( \mPe(k) \), are blue-tilted in nature. The maximum energy generated during inflation due to the coupling is proportional to the inflation energy scale, i.e. $\rho_{\rm EM} \propto H_I^4$
(for instance, see~\cite{Subramanian:2015lua, Haque:2020bip, Tripathy:2021sfb, Maiti:2025cbi}). We note that for \( n = 2 \), the spectrum of the magnetic field is scale-invariant, whereas for other values of \( n \), it becomes scale-dependent~\cite{Maiti:2025cbi}.  

Since we consider the coupling after inflation, there will be further production of the gauge field during the reheating era. Therefore, the reheating scenario must be constrained to prevent excessive production of the gauge field.

In models with a coupling of the form \( f^2 F_{\mu\nu} F^{\mu\nu} \), the effective electromagnetic gauge coupling is defined as \( \alpha_{\rm em} = e^2/f^2 \). As shown in Refs.~\cite{Sharma:2017eps, Maiti:2025cbi}, generating (nearly) scale-invariant magnetic spectra typically requires the coupling function \( f(\eta) \) to be small at the beginning of inflation and to grow toward unity by the end of inflation, thereby restoring the conformal symmetry of the gauge field action. This setup, however, implies a strong effective gauge coupling \( e^2/f^2 \) during the early stages of inflation, potentially leading to a strong coupling problem due to the large interaction strength with the background.

In contrast, in this work, we adopt a modified scenario where the coupling function begins with \( f(\eta) = 1 \), grows throughout inflation, and reaches a maximum at the end of inflation. To restore conformal symmetry, \( f(\eta) \) then decreases during reheating and returns to unity by the end of the reheating phase~\cite{Sharma:2017eps, Papanikolaou:2024cwr}. Because \( f(\eta) \geq 1 \) throughout the entire evolution in our model, the effective coupling strength \( e^2/f^2 \) remains bounded above by its standard value, avoiding the strong coupling issue.

To prevent the backreaction problem, we choose model parameters such that the total energy density of the generated electromagnetic field remains smaller than the background (radiation) energy density at the end of reheating. This approach, based on the modified evolution of the coupling function, allows us to address both the strong coupling and backreaction issues, while still producing magnetic fields of sufficient strength to match current large-scale observational constraints.

In the following subsection, we will solve the gauge field equation within generalized reheating scenarios. Before proceeding, however, we must first discuss the inflationary and reheating parameters that will be used throughout the later parts of this paper.

\paragraph{\underline{Defining the Inflationary and Reheating parameters:}}
Inflation is a phase during which the universe undergoes exponential expansion, and fluctuations on large scales are generated as modes exit the horizon. Generally, inflationary parameters such as its duration and the slow-roll parameters characterize the nature of inflation itself. However, since our primary objective is to study how the post-inflationary phase (the \textit{reheating era}) affects the subsequent evolution of the magnetic field, we provide a general parametrization of these quantities in terms of two key reheating parameters: the average equation-of-state parameter \( \wre \) and the reheating temperature \( \Tre \). This parametrization is valid for a wide class of perturbative reheating scenarios~\cite{PhysRevLett.113.041302}.

To begin, we fix the inflationary energy scale \( \HI \). Assuming that all scalar fluctuations originate from the inflaton field during inflation and that all tensor fluctuations at CMB scales stem from vacuum fluctuations, we can relate \( \HI \) to the amplitude of the scalar curvature power spectrum \( \As \) and the tensor-to-scalar ratio \( r \). The inflationary energy scale is then given by $\HI = \pi \Mp \sqrt{{r\,\As}/{2}}$.
In our analysis, two parameters are particularly relevant. The first is \( \ke \), the highest comoving mode that exits the horizon before the end of inflation. This effectively determines the duration of inflation after the CMB pivot scale \( \kpv \) leaves the horizon, and is related to the e-folds \( \NI \) via the relation \( \ke = \kpv\,\exp[\NI] \). The quantity \( \ke \) can be expressed as  \cite{Chakraborty:2024rgl, Maiti:2025cbi}
\begin{align}
    \ke = \left(\frac{43 \gre}{11}\right)^{1/3} \left(\frac{\pi^2 \gre}{90}\right)^{\sigma} \frac{\HI^{1 - 2 \sigma} \Tre^{4 \sigma - 1} T_0}{\Mp^{2 \sigma}},\label{eq:def_ke}
\end{align}
Similarly, another parameter relevant to our discussion is \( \kre \), which denotes the lowest comoving mode that re-enters the horizon before the end of reheating. This scale is determined by the reheating temperature \( \Tre \), and can be expressed as~\cite{Chakraborty:2024rgl, Maiti:2025cbi}
\begin{align}
    \kre \simeq 3.9 \times 10^6 \left(\frac{\Tre}{10^{-2}\, \text{GeV}}\right) \, \mathrm{Mpc}^{-1},
\end{align}
where \( \sigma = {1}/{3(1 + \wre)} \), and \( T_0 = 2.725 \, \text{K} \) is the present-day CMB temperature. We take \( a_0 = 1 \) as the present-day value of the scale factor. The reduced Planck mass is given by \( \Mp = 1/\sqrt{8\pi G} \simeq 2.14 \times 10^{18} \, \text{GeV} \), and \( \gre \simeq 106.7 \) denotes the effective number of relativistic degrees of freedom at the onset of the radiation-dominated era.

\subsection{Production during Inflation}

As this kind of coupling generates a non-helical magnetic field, where both the helicity modes `$+$' and `$-$' are equally enhanced from the vacuum, we can simplify our calculations by dropping the polarization index `\( \lambda \)` from Eq.~\eqref{eq:Ak_lambda}. We further define another variable \( \mAk(\eta) = f \Ak(\eta) \). Since we have already dropped the polarization index in this definition, we can rewrite the equation of motion (EoM) of the gauge field in terms of \( \mAk \), leading to the following equation~\cite{Martin:2007ue}
\begin{align}
    \mAk''(\eta) + \left(k^2 - \frac{f''}{f} \right) \mAk(\eta) = 0.
\end{align}

Replacing \( f''/f = n(n+1)/\eta^2 \), we obtain~\cite{Martin:2007ue}
\begin{align}
    \mAk''(\eta) + \left(k^2 - \frac{n(n+1)}{\eta^2} \right) \mAk(\eta) = 0.
\end{align}
The general solution of the above equation is given in terms of Bessel functions
\begin{align}
    \mAk(\eta) = \sqrt{-\eta} \left\{ c_1 \bJ_{n+\frac{1}{2}}(-k\eta) + c_2 \bY_{n+\frac{1}{2}}(-k\eta) \right\}.
\end{align}
To determine the values of \( c_1 \) and \( c_2 \), we impose the Bunch-Davies initial condition, assuming that all observed modes today were deep inside the Hubble horizon at the beginning of inflation. In the sub-horizon limit \( |-k\eta| \to \infty \), the solution of the gauge field reduces to a simple plane wave $\mAk(-k\eta \to \infty) \simeq {e^{-ik\eta}}/{\sqrt{2k}}$.
Using this as an initial condition, we obtain the integration constants \( c_1 \) and \( c_2 \) as~\cite{Subramanian:2015lua}
\begin{align}
    c_1 = \sqrt{\frac{\pi}{4k}} \frac{\exp(i\pi n/2)}{\cos(-\pi n)}, \hspace{1cm}
    c_2 = \sqrt{\frac{\pi}{4k}} \frac{\exp(i\pi(1-n)/2)}{\cos(-\pi n)}.
\end{align}
Combining these results, we express the solution of the gauge field during inflation in terms of the Hankel function as~\cite{Subramanian:2015lua, Maiti:2025cbi}
\begin{align}
    \mAk(\eta) = \sqrt{-\frac{\pi\eta}{4}} e^{i(n+1/2)\pi/2} \rmH^{(1)}_{n+\frac{1}{2}}(-k\eta).
\end{align}

\paragraph{\underline{Defined the Electric and Magnetic Spectral Energy Density at the End of Inflation:}\\}

Once we obtain the solution of the gauge field, utilizing this in the above Eqs.~\eqref{eq:def_pbe}, we can compute the magnetic and electric energy densities at the end of inflation, i.e., at $\eta = \ee$. We find~\cite{Maiti:2025cbi}
\begin{align}\label{eq14}
    \mathcal{P}_{B,\rm inf}(k,\eta) = \frac{d\rho_B}{d\ln k} = \frac{k^4}{8\pi a^4(\eta)} (-k\eta) \left|H^{(1)}_{n+\frac{1}{2}}(-k\eta)\right|^2 ,
\end{align}
\begin{align}\label{eq15}
    \mathcal{P}_{E,\rm inf}(k,\eta) = \frac{d\rho_E}{d\ln k} = \frac{k^4}{8\pi a^4(\eta)} (-k\eta) \left|H^{(1)}_{n-\frac{1}{2}}(-k\eta)\right|^2  .
\end{align}
Here, the subscript `$\mathrm{inf}$' denotes quantities defined during inflation. 

To analyze the spectral behavior of the magnetic and electric fields, we simplify the above equations. We are particularly interested in modes that are far outside the horizon before the end of inflation, i.e., $k < \ke$. In this regime, we take the super-horizon approximation $|-k\eta_{\rm end}| \ll 1$, allowing us to express the above equations in the following form~\cite{Subramanian:2015lua, Haque:2020bip, Tripathy:2021sfb}
\begin{align}\label{eq:pbi_1}
\left.  \begin{matrix}
    \mathcal{P}_{B,\rm inf}(k,\eta) = \frac{H_{\rm inf}^4}{8\pi} \frac{2^{2|n|+1} \Gamma^2\left(|n|+\frac{1}{2}\right)}{\pi^2} (-k\eta)^{-2|n|+4}, \\
    \mathcal{P}_{E,\rm inf}(k,\eta) = \frac{H_{\rm inf}^4}{8\pi} \frac{\Gamma^2\left(|n|-\frac{1}{2}\right) 2^{2|n|-1}}{\pi^2} (-k\eta)^{-2|n|+6}
  \end{matrix}\right\} \quad \text{for } n>\frac{1}{2},
\end{align}
\begin{align}\label{eq:pbi_2}
\left.  \begin{matrix}
    \mathcal{P}_{B,\rm inf}(k,\eta) = \frac{H_{\rm inf}^4}{8\pi} \frac{\Gamma^2\left(|n|-\frac{1}{2}\right) 2^{2|n|-1}}{\pi^2} (-k\eta)^{-2|n|+6}, \\
    \mathcal{P}_{E,\rm inf}(k,\eta) = \frac{H_{\rm inf}^4}{8\pi} \frac{2^{2|n|+1} \Gamma^2\left(|n|+\frac{1}{2}\right)}{\pi^2} (-k\eta)^{-2|n|+4}
  \end{matrix}\right\} \quad \text{for } n<-\frac{1}{2}.
\end{align}

From the above expressions, we see that for a coupling with $n > 0$, the magnetic field spectrum becomes scale-invariant for $n = 2$, while the electric field follows $\mathcal{P}_{E,\rm inf}(k) \propto k^2$. Since the electric field spectrum is blue-tilted, it avoids backreaction issues in this case. However, for $n > 2$, the magnetic spectrum becomes red-tilted, leading to an infrared (IR) divergence at large wavelengths. If we set the cutoff wavelength to the current CMB pivot scale, $k_{\rm cut} = k_* = 0.05~\Mpc^{-1}$, we can impose a strong constraint on $n$ such that the total energy density of the produced magnetic field remains below the total inflaton energy density. This gives an upper bound on the coupling parameter $n \leq 2.18$. This bound is derived assuming an inflationary duration corresponding to $\NI = 55$ e-folds, where $\NI$ represents the number of e-folds during inflation. 

Additionally, a scale-invariant magnetic field is also possible for $n = -3$. However, as our analysis aims to simultaneously avoid both the backreaction and strong coupling problems, we discard the $n < 0$ cases due to their inherent strong coupling issues.
\color{black}
\subsection{Production during Reheating}

As discussed earlier, the coupling function \( f(\eta) \) remains active during reheating, leading to additional magnetic field generation. Incorporating its functional form into the equation of motion (EoM) for \( \mAk \) during reheating, we obtain  
\begin{align}
    \mAk''+\left(k^2-\frac{\alpha(\alpha+1)}{\eta^2}\right)\mAk=0.
\end{align}  
Here, \( \alpha \) is defined as  
\begin{align}
    \alpha = \frac{2m}{1+3\wre} = \frac{2n\beta}{1+3\wre},
\end{align}  
where \( \beta \) depends on the reheating dynamics, influencing the spectral behavior of the magnetic field.  

The general solution to this equation is  
\begin{align}
    \mAk^{\mathrm{re}}(k,\eta>\ee) = \sqrt{k\eta} \left\{ d_1 J_{\alpha+1/2}(k\eta) + d_2 J_{-\alpha-1/2}(k\eta) \right\},
\end{align}  
where \( J_{\nu} \) are Bessel functions, and \( d_1 \) and \( d_2 \) are integration constants. These constants are determined by enforcing the continuity of \( \mAk \) and its first derivative at the transition point \( \eta = \ee \).  

Applying these conditions yields  
\begin{subequations}
    \begin{align}
        d_1 &= \frac{\pi \xe^{1/2}}{2\cos(\pi\alpha)} \left[ \mAki(\xe) \tilde{\mJ}_{-\alpha-1/2}(\xe) + {\mAki}'(\xe) \mJ_{-\alpha-1/2}(\xe) \right], \\
        d_2 &= \frac{\pi \xe^{1/2}}{2\cos(\pi\alpha)} \left[ \mAki(\xe) \tilde{\mJ}_{\alpha+1/2}(\xe) - {\mAki}'(\xe) \mJ_{\alpha+1/2}(\xe) \right],
    \end{align}
\end{subequations}  
where \( \xe = -k\ee = k/\ke \), and the auxiliary functions are  
\begin{subequations}
   \begin{align}
    \tilde{\mJ}_{\alpha+1/2}(\xe) &= \frac{(1+\alpha)\mJ_{\alpha+1/2}(\xe) - \xe \mJ_{\alpha+3/2}(\xe)}{\xe}, \\
    \tilde{\mJ}_{-\alpha-1/2}(\xe) &= \frac{\alpha\mJ_{-\alpha-1/2}(\xe) + \xe \mJ_{-\alpha+1/2}(\xe)}{\xe}.
   \end{align}
\end{subequations}  
Since we are primarily interested in modes that exited the horizon well before the end of inflation, i.e., \( k < \ke \), we focus on the regime where \( \xe \ll 1 \).
  
 \begin{figure*}
\includegraphics[width=0.45\linewidth]{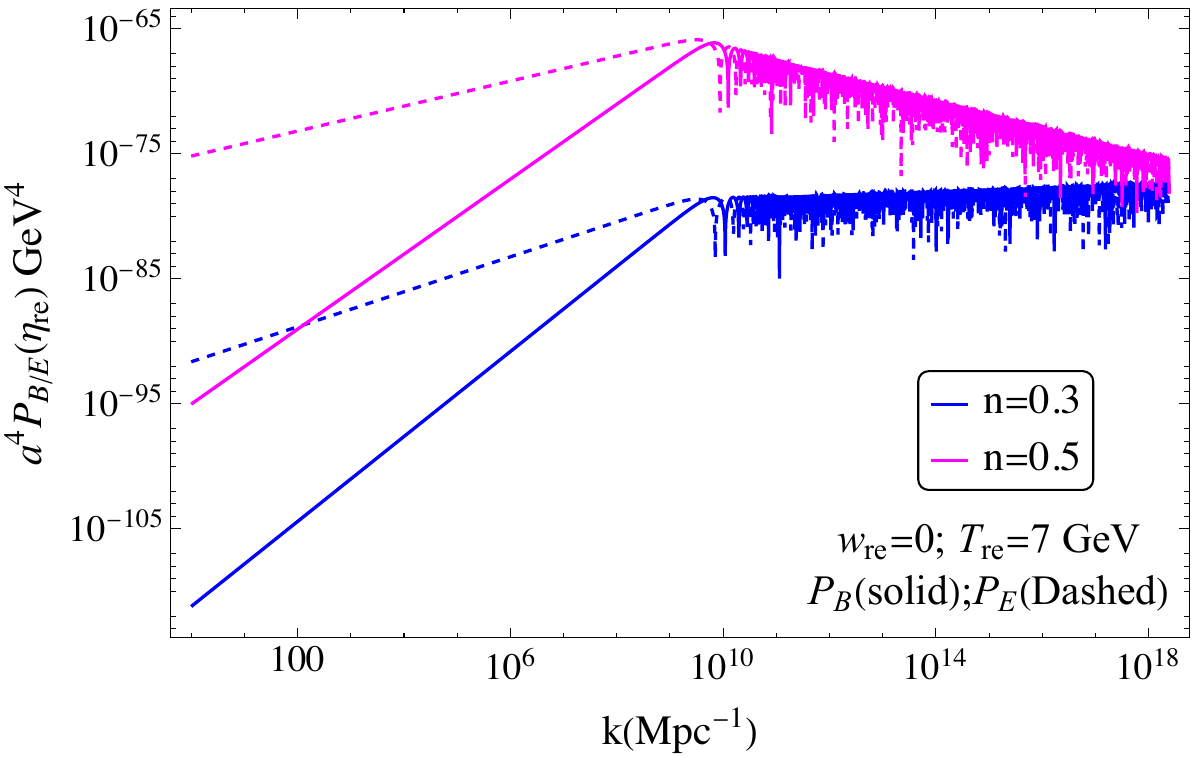}
\includegraphics[width=0.45\linewidth]{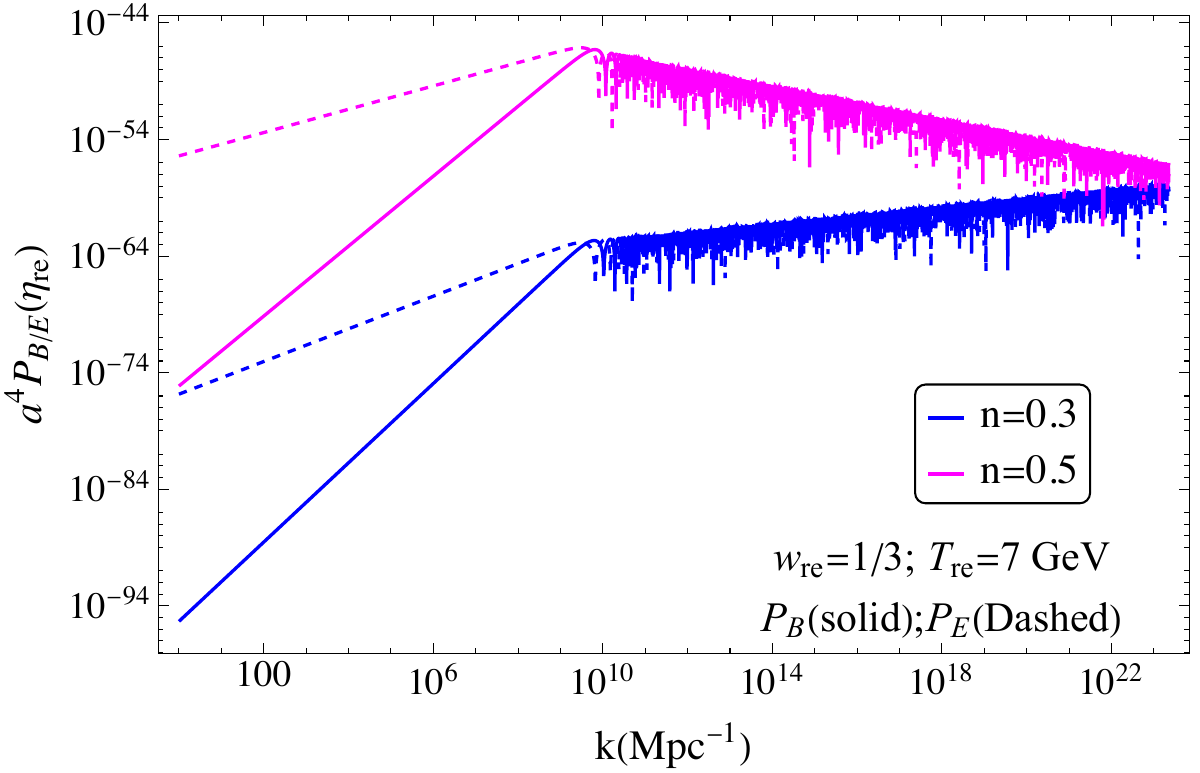}
\caption{Comoving electric and magnetic power spectra, \( a^4\mPbe(k) \), as a function of the comoving wavenumber \( k \) (in \( \Mpc^{-1} \)). \textbf{Left panel:} Reheating scenario with an equation of state \( \wre = 0.0 \). \textbf{Right panel:} Reheating scenario with \( \wre = 1/3 \). In both panels, the blue lines correspond to \( n = 0.3 \), while the magenta lines represent \( n = 0.5 \). The reheating temperature is fixed at \( \Tre = 7\,\Gev \). Solid lines indicate the comoving magnetic spectral energy density, whereas dashed lines represent the comoving electric spectral energy density.}
    \label{fig:Pbe_vs_k_nb}
\end{figure*}
 \begin{figure*}
\includegraphics[width=0.45\linewidth]{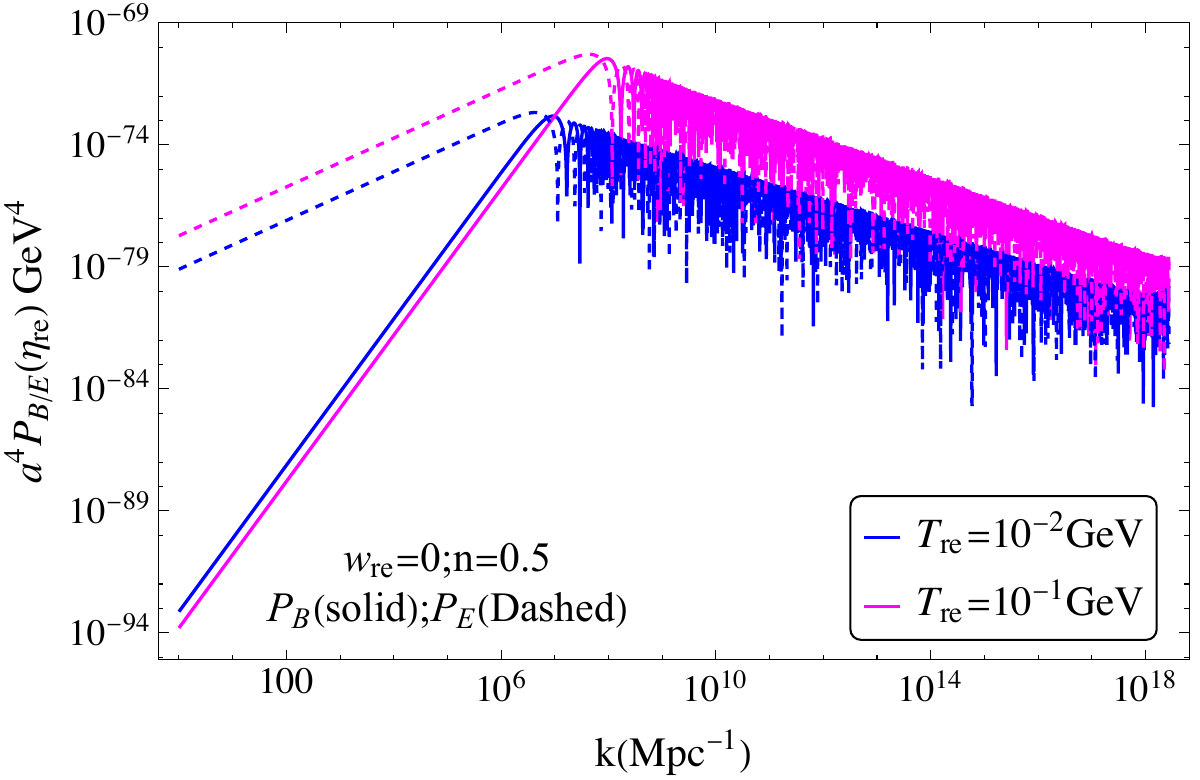}
\includegraphics[width=0.45\linewidth]{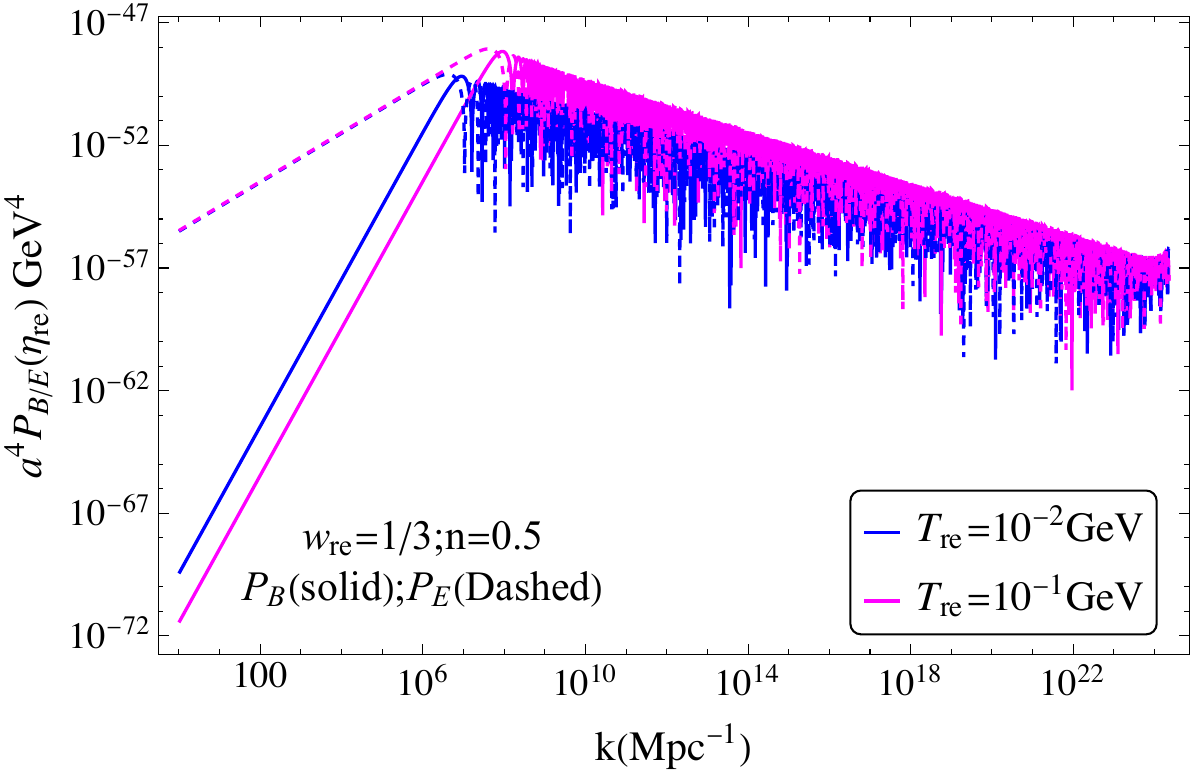}
\caption{Comoving electric and magnetic spectral energy density, \( a^4 \mPbe \), as a function of the comoving wavenumber \( k \) for two different reheating scenarios. \textbf{Left panel:} Equation of state \( \wre = 0.0 \) with \( n = 0.5 \). \textbf{Right panel:} Equation of state \( \wre = 1/3 \) with \( n = 0.5 \). In both panels, blue lines correspond to \( \Tre = 10^{-2} \,\Gev \), while magenta lines represent \( \Tre = 10^{-1} \,\Gev \). Solid lines indicate the magnetic field, whereas dashed lines correspond to the electric field.} 
\label{fig:pbe_vs_k_tre}
\end{figure*}

Under the super-horizon approximation, the expressions for the coefficients \( d_1 \) and \( d_2 \) simplify to  
\begin{subequations}
    \begin{align}
        d_1 &\simeq \sqrt{\frac{\pi}{4k}} \left(-\xe^{-n-\alpha-1} \frac{i 2^{n+\alpha}(\alpha-n)\Gamma(n+1/2) }{\cos(\pi \alpha)\Gamma(1/2-\alpha)}+\mathcal{O}(2)\right),\\
        d_2 &\simeq \sqrt{\frac{\pi}{4k}} \left(-\xe^{-n+\alpha} \frac{i 2^{n-\alpha-1}(1+n+\alpha)\Gamma(n+1/2)}{\cos(\pi \alpha)\Gamma(\alpha+3/2)}+\mathcal{O}(2)\right).
    \end{align}
\end{subequations}

To determine the spectral energy densities of the magnetic and electric fields, we substitute the mode function \( \mAk \) into the relevant expressions. However, further simplifications are required to analyze the spectral behavior effectively.

Since we define the comoving spectrum at the end of reheating (\(\eta = \ere\)), the spectrum can be categorized into two distinct regimes:  
\begin{itemize}  
    \item \textbf{Super-horizon limit:} \( k < \kre \)  
    \item \textbf{Sub-horizon limit:} \( \kre < k < \ke \)  
\end{itemize}  

\subsection*{Super-Horizon Limit (\( k_* < k < \kre \))}  
For super-horizon modes, the spectral energy densities of the magnetic and electric fields are given by  
\begin{subequations}\label{eq:pb_sup_h}
    \begin{align}
        \pb(k,\ere)
        &\simeq \frac{\HI^4}{8\pi} \frac{2^{2n-1}(\alpha-n)^2\Gamma^2(n+1/2)}{\cos^2(\pi\alpha)\Gamma^2(1/2-\alpha)\Gamma^2(\alpha+3/2)} \left( \frac{\xre}{\xe} \right)^{2(\alpha+1)} \left( \frac{k}{\ke} \right)^{4-2n} \left( \frac{\ae}{\are} \right)^4,\label{eq:pb_suer} \\
        \pe(k,\ere)
        &\simeq \frac{\HI^4}{8\pi} \frac{2^{2n+1}(\alpha-n)^2\Gamma^2(n+1/2)}{\cos^2(\pi\alpha)\Gamma^2(1/2-\alpha)\Gamma^2(\alpha+1/2)} \left( \frac{\xre}{\xe} \right)^{2\alpha} \left( \frac{k}{\ke} \right)^{2-2n} \left( \frac{\ae}{\are} \right)^4.\label{eq:pe_super}
    \end{align}
\end{subequations}  
The factor \( (\ae/\are)^4 \) arises due to the expansion of the universe from the end of inflation to the completion of reheating. Notably, the presence of the coupling function during reheating enhances the total energy density of super-horizon modes by a factor of \( (\xre/\xe)^{2(\alpha+1)} \). While the spectral indices of the magnetic and electric fields are determined by the parameter \( n \), the post-inflationary coupling governs the amplification of these fields.

A particularly interesting case arises for \( n=2 \), where the magnetic field spectrum remains scale-invariant, while the electric field spectrum exhibits a red tilt. This red tilt leads to an ultraviolet (UV) divergence in the electric field spectrum. However, even in this case, a truly scale-invariant magnetic spectrum cannot be achieved due to backreaction issues, which will be discussed later.

\subsection*{Sub-Horizon Limit (\( \kre < k < \ke \))}  
For sub-horizon modes, the spectral energy densities are given by  
\begin{subequations}\label{eq:pb_sub_h}
    \begin{align}
        \pb(k,\ere)
        &\simeq \frac{\HI^4}{8\pi} \frac{2}{\pi} \frac{2^{2(n+\alpha)}(\alpha-n)^2\Gamma^2(n+1/2)}{\cos^2(\pi\alpha)\Gamma^2(1/2-\alpha)} \left( \frac{k}{\ke} \right)^{2-2(n+\alpha)} \left( \frac{\ae}{\are} \right)^4, \\
        \pe(k,\ere)
        &\simeq \frac{\HI^4}{4\pi} \frac{2^{2(n+\alpha)}(\alpha-n)^2\Gamma^2(n+1/2)}{\cos^2(\pi\alpha)\Gamma^2(1/2-\alpha)} \left( \frac{k}{\ke} \right)^{2-2(n+\alpha)} \left( \frac{\ae}{\are} \right)^4.
    \end{align}
\end{subequations}  
In this regime, both the magnetic and electric field spectra share the same spectral index, which is consistently red-tilted. As a result, the total electromagnetic (EM) energy density attains its maximum at \( k = \kre \).
  
Generally, super-horizon modes exhibit a blue-tilted spectrum, whereas sub-horizon modes remain red-tilted. This spectral behavior arises from the specific form of the gauge field coupling function, which is chosen to address both the backreaction and strong coupling problems simultaneously. Moreover, the coupling function is designed to restore the conformal nature of the gauge field by reaching unity at the end of reheating.

To analyze the spectral behavior of the magnetic and electric spectral energy densities, denoted as \( \mPb \) and \( \mPe \), we have plotted the comoving magnetic and electric spectral energy densities, defined as \( a^4\mathcal{P}_{\rm B/E} \), at the end of reheating as a function of the comoving wavenumber (in \( \Mpc^{-1} \)).  

In Fig.~\ref{fig:Pbe_vs_k_nb}, we consider two different values of the coupling constant: \( n=0.3 \) (blue) and \( n=0.5 \) (red), for two different equations of state (EoS): \( \wre=0.0 \) (left panel) and \( \wre=1/3 \) (right panel). In both cases, the reheating temperature is fixed at \( \Tre = 7\,\Gev \). The solid lines represent the magnetic spectral energy density (\(\mPb\)), whereas the dashed lines correspond to the electric spectral energy density (\(\mPe\)).  

For super-horizon modes (\( k<\kre \)), the spectral behavior of the electric and magnetic fields differs, as predicted earlier. However, for sub-horizon modes (\( k>\kre \)), both spectral energy densities follow the same behavior, as discussed in Eqs.~\eqref{eq:pb_sub_h}. Additionally, depending on the coupling nature, if we aim to generate a significant magnetic field on large scales, we find that in most cases, the maximum energy density is stored at \( k\simeq\kre \) for both magnetic and electric fields.  

Similarly, in Fig.~\ref{fig:pbe_vs_k_tre}, we plot the comoving spectral energy density \( a^4\mathcal{P}_{\rm B/E} \) as a function of the comoving wavenumber \( k \) for a fixed coupling parameter \( n=0.5 \), considering two different EoS values: \( \wre=0.0 \) (left panel) and \( \wre=1/3 \) (right panel). In both figures, blue represents \( \Tre = 10^2\,\Gev \), while red corresponds to \( \Tre = 0.1\,\Gev \). Here, the dashed lines indicate the comoving spectral energy density of the electric field, whereas the solid lines correspond to the comoving spectral energy density of the magnetic field.  

From Eqs.~\eqref{eq:pb_sup_h}, we observe that the super-horizon scaling behavior of the magnetic and electric fields is dictated by the coupling parameter \( n \). Since we consider a single value, \( n=0.5 \), the spectral tilt remains the same in both panels, even for sub-horizon modes at a fixed reheating temperature.  

Furthermore, in the case of \( \wre = 1/3 \), the spectral energy density of the super-horizon electric field modes retains the same amplitude, irrespective of the reheating temperature. This seemingly counterintuitive behavior arises due to the factor \( (\xre/\xe)^{2\alpha} \). For \( \wre = 1/3 \), the evolution of the relevant background quantities compensates in such a way that the electric field undergoes the same enhancement regardless of the reheating temperature.

However, for the magnetic field, we find that even in the \( \wre=1/3 \) reheating scenario, an additional enhancement factor, \( (\ke/\kre)^2 \), exists. As a result, changing the reheating temperature alters the amplitude of the magnetic field due to this factor, while the spectral index remains unchanged.

\paragraph{\underline{ Present-day Magnetic Field Strength}}

In this type of magnetogenesis model, the electromagnetic fields restore their conformal symmetry after reheating as the coupling function becomes \( f(\eta > \ere) = 1 \). Since there are no additional dynamics that break the conformal nature of the electromagnetic (EM) fields, the effective production of gauge fields ceases after reheating.  

During reheating, all fundamental particles are already produced, and the universe reaches a high-temperature state, behaving as a highly dense conducting plasma. As a result, the electric field component rapidly decays and is washed out from the universe. Therefore, after reheating (during the radiation-dominated era), the only surviving component is the magnetic field.  

Since we are not considering magnetohydrodynamics (MHD) effects and are interested in modes that remain far outside the horizon during the radiation-dominated era, we can safely ignore MHD effects for those modes. Typically, these large-scale modes evolve adiabatically after their production, with the energy density of the magnetic field scaling as \( \rho_{\rm B} \propto a^{-4} \) due to the expansion of the universe.  

Under this assumption, the present-day magnetic field strength can be expressed as  
\begin{align}
    \textbf{B}_0(k) \simeq \frac{\ke^2}{(8\pi)^{1/2}}  
    \frac{2^{n-1/2}(\alpha-n)\Gamma(n+1/2)}
    {\cos(\pi\alpha)\Gamma(1/2-\alpha)\Gamma(\alpha+3/2)} 
    \left( \frac{\ke}{\kre} \right)^{\alpha+1}
    \left( \frac{k}{\ke} \right)^{2-n}.
\end{align}
 \begin{figure*}
\includegraphics[width=0.45\linewidth]{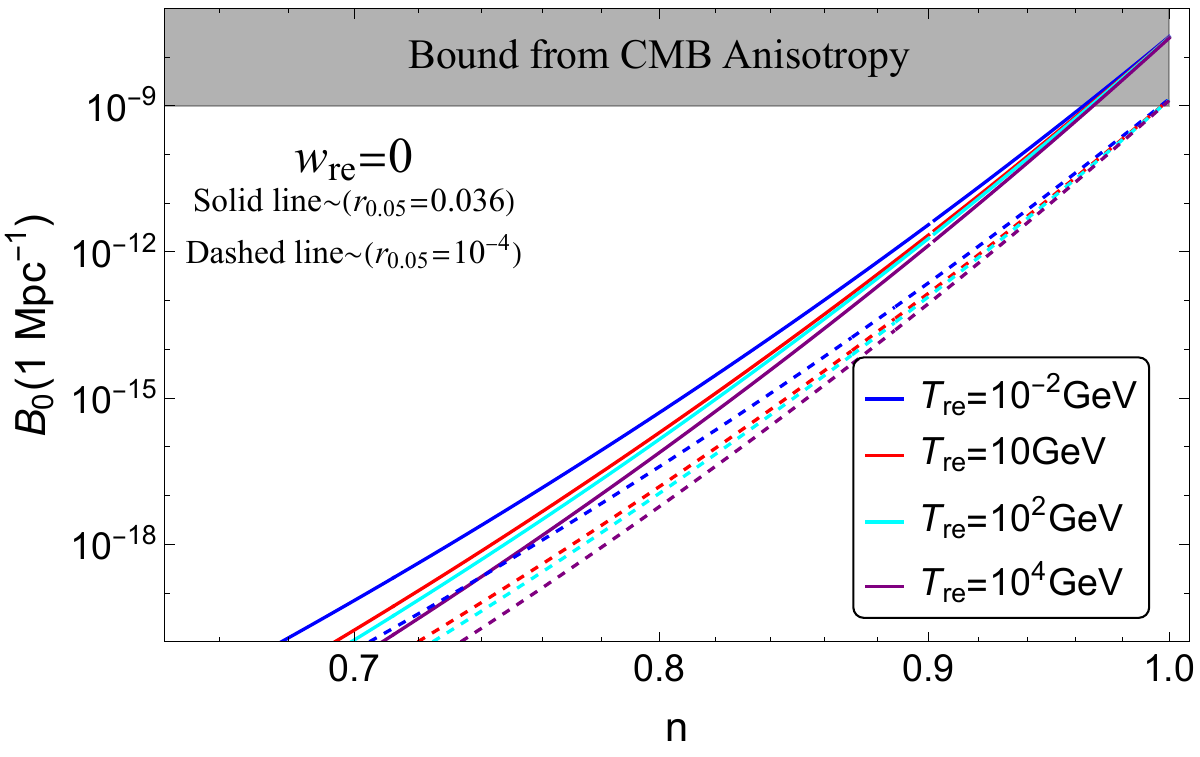}
\includegraphics[width=0.45\linewidth]{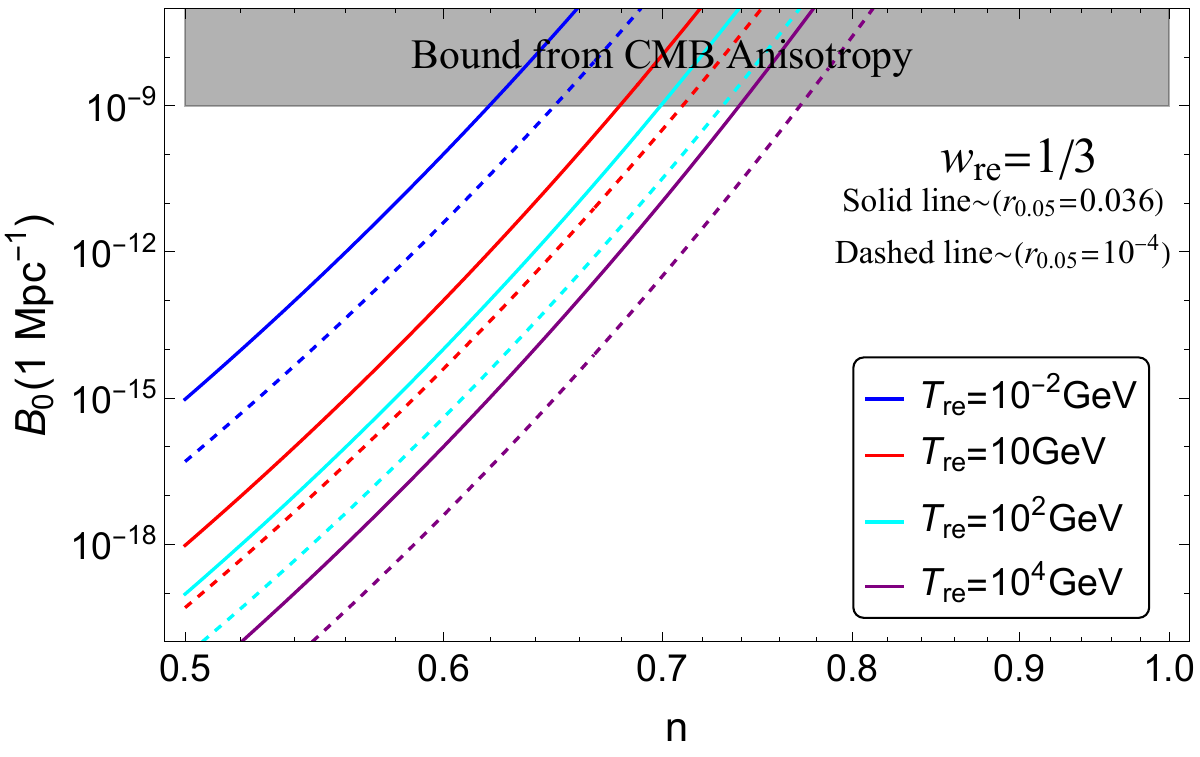}
\caption{Present-day magnetic field strength \( B_0 \) as a function of the coupling constant \( n \) at the \( 1~\text{Mpc}^{-1} \) scale. \textbf{Left panel:} Equation of state \( \wre = 0 \). \textbf{Right panel:} Equation of state \( \wre = 1/3 \) for reheating scenarios. In both figures, different colors represent different reheating temperatures \( (\Tre) \). The solid lines correspond to the tensor-to-scalar ratio \( r_{0.05} = 0.036 \), while the dashed lines represent \( r_{0.05} = 10^{-4} \).} 
\label{fig:b0_vs_n}
\end{figure*}

\begin{center}
\begin{table}[t]
\centering
    \begin{tabular}{c ||c c c c|| c c c c  }
    \hline
    \hline
    EoS & \multicolumn{4}{c||}{$\wre=0.0$} & \multicolumn{4}{c}{$\wre=1/3$} \\
    \hline
    $n$  & 0.6 & 0.7 & 0.8 & 0.9  & 0.4 & 0.45 & 0.5 & 0.55\\
    $\Tre$(GeV) &  $2.54\times 10^6$ & $10^4$ & $ 24.5$ & $0.043$  & $5.31\times 10^7$ & $ 1.02\times 10^5$ & $ 58.4$ & $ 10^{-2}$\\
    $\textbf{B}_0$ (G) & $5.5 \times 10^{-26}$ & $4.3\times 10^{-21}$ & $1.7\times 10^{-16}$ & $3.3\times 10^{-12}$ &  $1.7\times 10^{-30}$ & $2.9\times 10^{-25}$ & $1.66\times 10^{-19}$ & $3.2\times 10^{-13}$\\
    \hline
    \hline
    
    \end{tabular}
         \caption{In the above table, we have estimated the upper bound of reheating temperature to prevent the backreaction of the gauge field with the background for a specific set of coupling constant $n$ for two different EoS $\wre=0$ and $\wre=1/3$. here we also listed the corresponding present-day magnetic field strength $\textbf{B}_0$ ( in Gauss unit) for these specific parameters.}
    \label{tab:param_est}
 \end{table}
\end{center}
To analyze the dependency of the present-day magnetic field strength \( \textbf{B}_0(k) \) on the magnetogenesis parameters (such as the initial coupling constant \( n \)) and the reheating dynamics (in terms of reheating temperature \( \Tre \) and average equation of state \( \wre \)), we have plotted \( \textbf{B}_0(1~\text{Mpc}^{-1}) \) as a function of the coupling parameter \( n \) in Fig.~(\ref{fig:b0_vs_n}).  

In the left panel of Fig.~\ref{fig:b0_vs_n}, we show the present-day magnetic field strength evaluated at the \( 1~\Mpc \) scale for \( \wre = 0 \), where different colors correspond to different values of the reheating temperature \( \Tre \). In the right panel, we plot \( B_0(1~\Mpc^{-1}) \) as a function of the coupling parameter \( n \) for radiation-like reheating scenarios with \( \wre = 1/3 \).  

In both figures, we consider two different values of the tensor-to-scalar ratio \( r_{0.05} \): solid lines correspond to \( r_{0.05} = 0.036 \), while dashed lines represent \( r_{0.05} = 10^{-4} \). Since we have fixed the Hubble parameter during inflation as $\HI = \pi\Mp\sqrt{{r_{0.05} A_s}/{2}}$,
changing the value of \( r_{0.05} \) effectively modifies the inflationary energy scale.  

From both figures, we observe that for a fixed equation of state \( \wre \) and reheating temperature \( \Tre \), an increase in the coupling parameter \( n \) leads to a gradual increase in the present-day magnetic field strength. This behavior arises due to the initial value of the gauge field generated during inflation, which is governed by the factor \( (k/\ke)^{-2|n|+4} \).  

Additionally, we find that lowering the reheating temperature enhances the overall strength of the gauge field. This effect is due to the prolonged duration of the reheating period, which allows for increased production of the gauge field during reheating.  

\paragraph{\underline{Backreaction issue}}\

Including the coupling function \( f(\eta) \) in the action with the electromagnetic (EM) field breaks conformal symmetry, leading to significant EM field generation during both inflation and reheating. To avoid backreaction with the inflaton field at the end of reheating, we must compute the total energy density produced due to the coupling. The total EM energy density at the end of reheating is defined as
\begin{align}
    \rho_{\rm em}(\ere) = \int_{\kre}^{\ke} d\ln(k) \left\{\mPb(k,\ere) + \mPe(k,\ere)\right\}, \label{eq:rho_em}
\end{align}
where the integration limits are chosen from \( \kre \) to \( \ke \), since \( \kre \) is the highest mode that can re-enter the horizon at the end of reheating. Notably, since the spectrum peaks at \( k \simeq \kre \), extending the lower limit to \( \kcmb \) does not significantly alter the result.

The fractional energy density of the EM field at the end of reheating is then defined as $\delta_{\rm em} = \frac{\rho_{\rm em}(\ere)}{\rho_{\rm c}(\ere)}$,
where \( \rho_{\rm c}(\ere) = 3H^2(\ere) \Mp^2 \), and \( H(\ere) \) is the Hubble parameter at the end of reheating. Utilizing Eq.~\eqref{eq:pb_sub_h} in Eq.~\eqref{eq:rho_em}, we obtain:
\begin{align}
    \delta_{\rm em}(\ere) &= \frac{1}{24} \left( \frac{\HI}{\mpl} \right)^2 
    \left( \frac{\kre}{\ke} \right)^{2 - 2(n+\alpha)} 
    \frac{2^{2n+1} (\alpha - n)^2 \Gamma^2(n+1/2)}{\cos^2(\alpha\pi) \Gamma^2(1/2-\alpha)} \notag \\
    &\quad \times \left\{ \frac{1}{\Gamma^2(\alpha+1/2)} + \frac{2^{2\alpha}}{\pi} \right\} 
    \left( \frac{\ae}{\are} \right)^{1 - 3\wre}.
\end{align}
Here, the term \( (\ae/\are)^{1-3\wre} \) accounts for the relative dilution of the EM field energy density compared to the background.

\begin{figure*}
\includegraphics[width=0.45\linewidth]{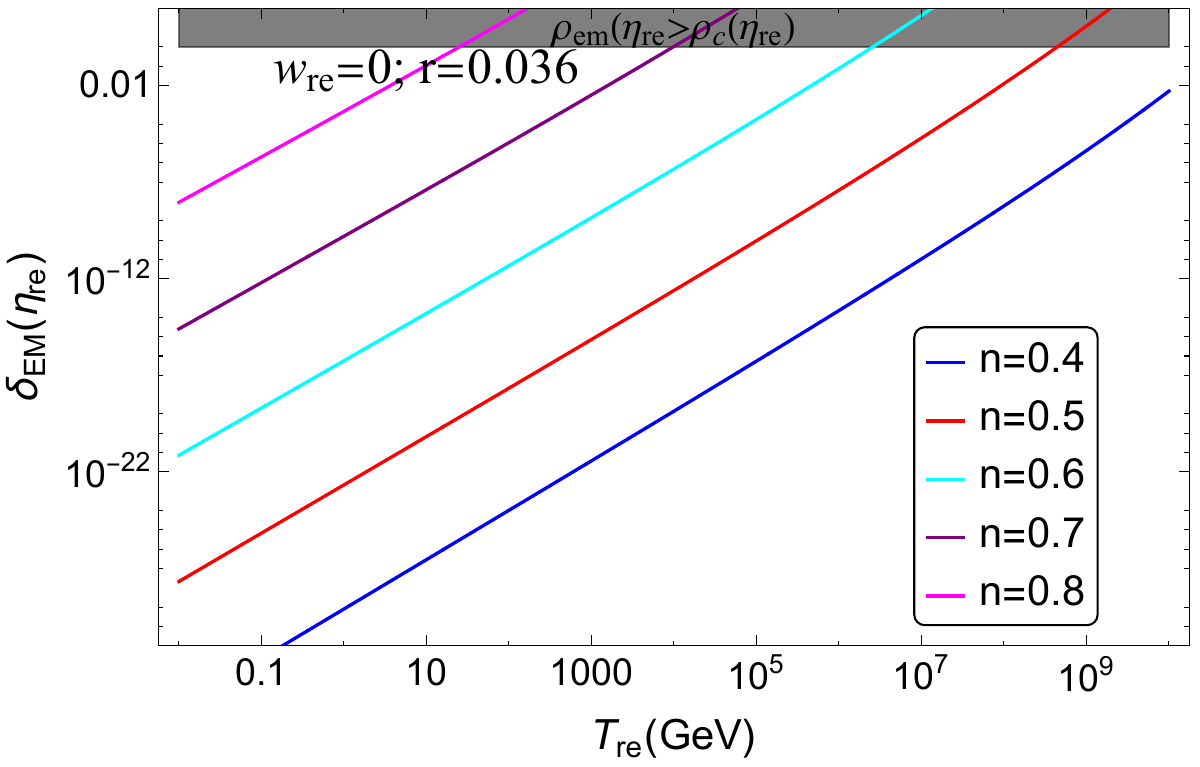}
\includegraphics[width=0.45\linewidth]{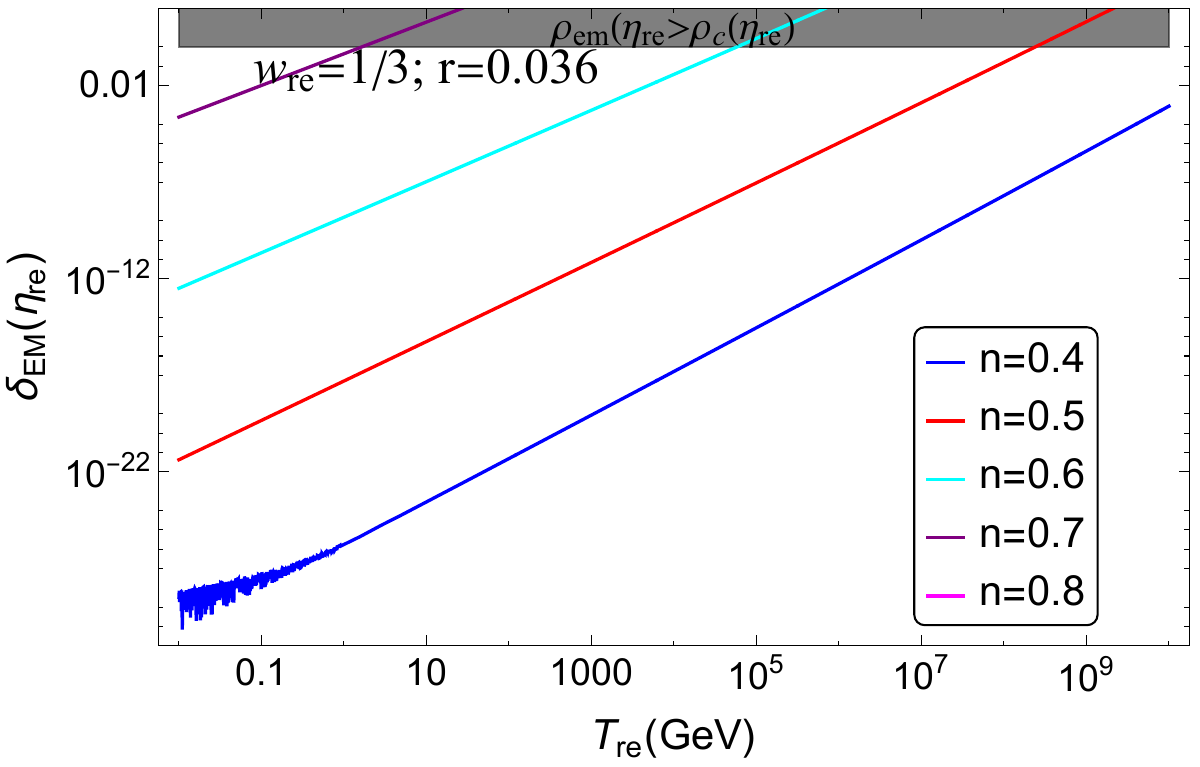}
\caption{Fractional energy density of the EM field, \( \delta_{\rm EM} \), as a function of reheating temperature \( \Tre \) (in GeV) for two different values of the equation of state: \( \wre = 0.0 \) (left) and \( \wre = 1/3 \) (right). In both figures, different colored lines correspond to different values of the coupling parameter \( n \). The deep gray shaded region represents the parameter space where backreaction effects become significant.} 
\label{fig:delta_tre}
\end{figure*}

In Fig.~\ref{fig:delta_tre}, we illustrate how the fractional energy density of the produced EM field depends on reheating scenarios and coupling parameters. Specifically, in Fig.~\ref{fig:delta_tre}, we plot \( \delta_{\rm em}(\ere) \) as a function of reheating temperature \( \Tre \) (in GeV) for two distinct equations of state: \( \wre = 0.0 \) (left) and \( \wre = 1/3 \) (right). Different colors represent different values of the coupling parameter \( n \). 

From the figure, we observe that for a fixed reheating temperature (e.g., \( \Tre = 0.1\, \text{GeV} \)), increasing \( n \) leads to a gradual increase in the total energy density. This arises because the initial EM energy density produced during inflation depends on \( n \). A larger \( n \) results in a less blue-tilted spectrum, leading to a comparatively higher energy density on larger scales.

Since the coupling function during reheating is no longer a free parameter, it strongly depends on reheating dynamics. This effect is evident in both figures: when the EoS changes, the EM field production rate is significantly affected due to background evolution. For a matter-like evolution (\( \rho_\phi \propto a^{-3} \)), the universe spends a longer duration in reheating compared to a radiation-like scenario (\( \rho_\phi \propto a^{-4} \)). 

To satisfy the condition \( f(\ere) = 1 \), for \( \wre = 0 \) reheating, we must effectively choose a lower value of \( \alpha \) for a fixed \( n \), while for \( \wre = 1/3 \), \( \alpha \) is relatively larger. A higher \( \alpha \) corresponds to stronger EM field production during reheating. Consequently, for a fixed \( n \) and \( \Tre \), radiation-like evolution results in greater EM field production.

\begin{figure*}
\includegraphics[width=0.45\linewidth]{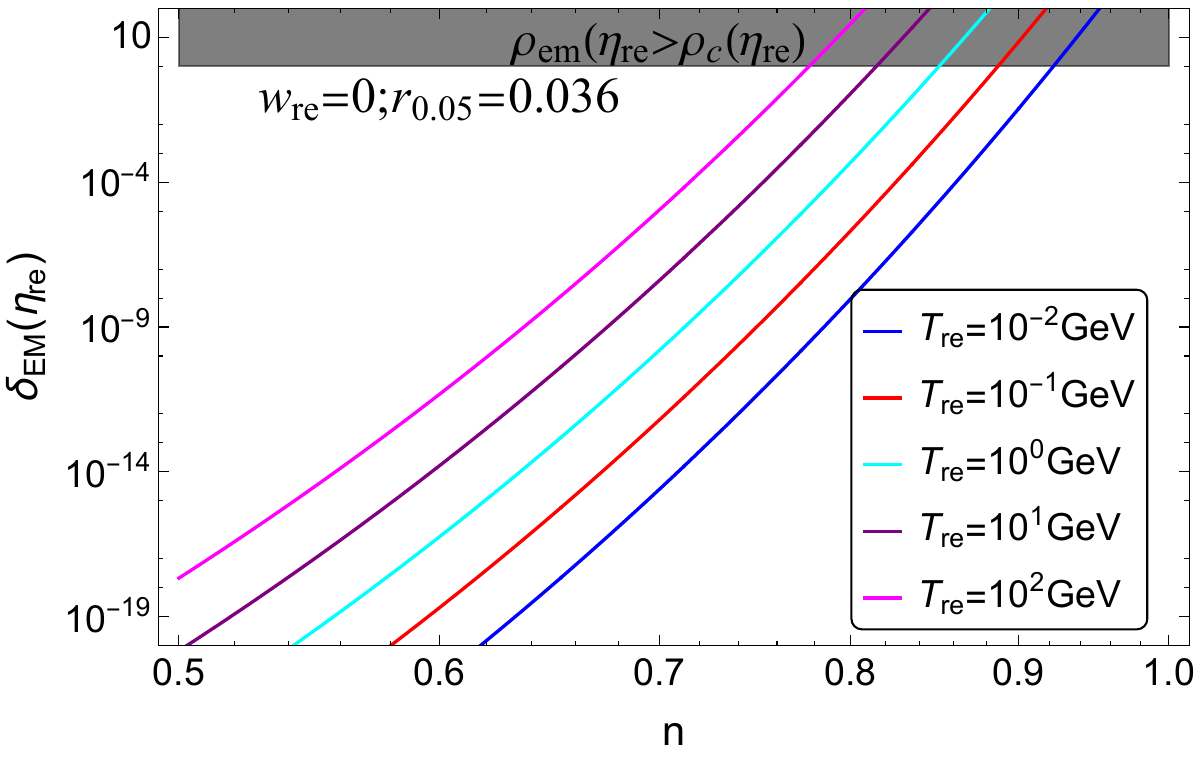}
\includegraphics[width=0.45\linewidth]{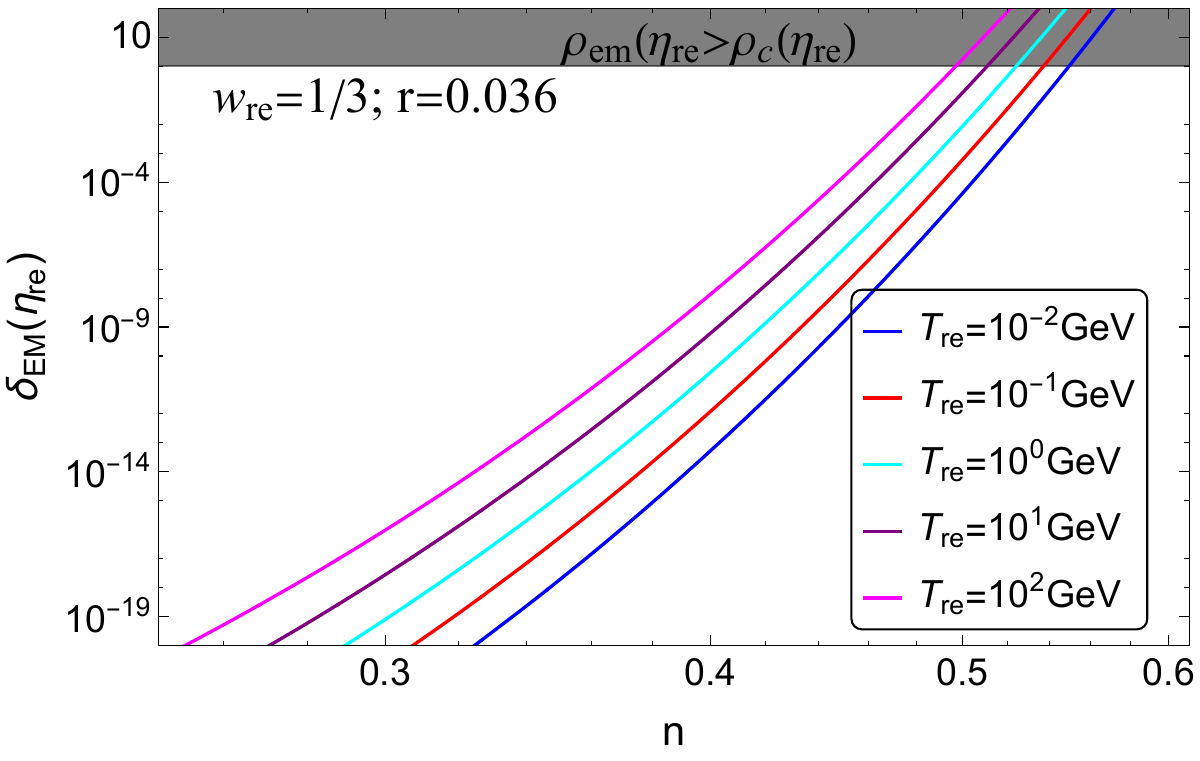}
\caption{Fractional energy density \( \delta_{\rm EM} \) as a function of the coupling parameter \( n \) for two different reheating scenarios: \( \wre = 0.0 \) (left) and \( \wre = 1/3 \) (right). In both figures, different colored lines represent different values of the reheating temperature \( \Tre \) (in GeV). The deep gray shaded region indicates the parameter space where backreaction effects become significant.} 
\label{fig:delta_n}
\end{figure*}

Similarly, in Fig.~\ref{fig:delta_n}, we plot \( \delta_{\rm em}(\ere) \) as a function of \( n \) for \( \wre = 0.0 \) (left) and \( \wre = 1/3 \) (right), where different colors indicate different reheating temperatures. We find that to avoid backreaction for \( \wre = 0.0 \), the maximum allowed coupling parameter is \( n < 1 \) for a wide range of reheating temperatures. For example, for \( \Tre = 1\,\text{GeV} \) with \( \wre = 0.0 \), the maximum allowed value is \( n^{\rm max} \simeq 0.83 \), while for \( \wre = 1/3 \), it is \( n^{\rm max} \simeq 0.53 \).

Table~\ref{tab:param_est} presents estimates of the lowest possible reheating temperature \( \Tre \) (in GeV) for \( \wre = 0.0 \) and \( \wre = 1/3 \), along with the corresponding present-day magnetic field strength at \( 1\, \Mpc \). We find that:
- For \( \wre = 0 \), ensuring a backreaction-free scenario requires \( 0.75 \leq n \leq 0.92 \), while the reheating temperature remains low: \( 10^{-2} \leq \Tre \leq 10^2 \) GeV.
- For \( \wre = 1/3 \), the allowed \( n \) range is narrower: \( 0.5 \leq n \leq 0.55 \), with a similar reheating temperature range.

Interestingly, a higher reheating temperature leads to stronger magnetic field generation on small scales. However, the present-day magnetic field strength at \( 1\,\Mpc \) remains relatively small.

\section{Generation of Gravitational Waves}\label{sec:GWs}

In the previous section, we discussed how the additional nontrivial coupling with the gauge field can generate large-scale magnetic fields. The choice of the coupling function allows for the generation of a magnetic field even after inflation. During reheating, the production of the gauge field is significantly enhanced, to the extent that it can backreact on the inflaton field. Given this strong gauge field production, it is expected that it can also act as a source of gravitational waves (see, for instance,~\cite{Sorbo:2011rz, Caprini:2014mja, Ito:2016fqp, Sharma:2019jtb, Okano:2020uyr, Maiti:2025cbi, Maiti:2025rkn, Maiti:2024nhv}).

In the presence of an electromagnetic field, the anisotropic stress component of the energy-momentum tensor can source tensor perturbations, denoted by $h_{ij}$. The equation of motion governing the evolution of $h_{ij}$ is given by~\cite{Sorbo:2011rz, Caprini:2014mja, Ito:2016fqp, Sharma:2019jtb, Okano:2020uyr, Maiti:2025cbi, Maiti:2025rkn, Maiti:2024nhv}
\begin{align}\label{eq4}
     h_{ij}''(\eta,\mathbf{x})+2\mathcal{H} h_{ij}'(\eta,\mathbf{x})-\nabla^2 h_{ij}(\eta,\mathbf{x})=\frac{2}{M_{\text{Pl}}^2}\mathcal{P}^{lm}_{ij} T_{lm}(\eta, \mathbf{x}),
\end{align}
where $h_{ij}$ is a traceless tensor, satisfying $\partial^i h_{ij} = h^i_i = 0$. The quantity $\mathcal{H} = a'(\eta)/a(\eta)$ represents the conformal Hubble parameter. The term $\mathcal{P}^{lm}_{ij}$ is the transverse traceless projector, given by $\mathcal{P}_{ij}^{lm} = P^l_i P^m_j - P_{ij} P^{lm}/2$, where $P_{ij} = \delta_{ij} - \partial_i \partial_j/\Delta$. The term $T_{lm}$ represents the spatial part of the energy-momentum tensor of the gauge field.

The tensor perturbations $h_{ij}(\eta,\mathbf{x})$ evolving in the Friedmann universe can be decomposed in terms of their Fourier modes, $h_{\mathbf{k}}^{\lambda}(\eta)$, as follows~\cite{Maiti:2024nhv}
\begin{equation}
    h_{ij}(\eta,\mathbf{x}) = \sum_{\lambda=(+,\times)} \int \frac{\mathrm{d}^3\mathbf{k}}{(2\pi)^{3/2}} e_{ij}^{\lambda}(\mathbf{k}) h_{\mathbf{k}}^{\lambda}(\eta) \mathrm{e}^{i\mathbf{k} \cdot \mathbf{x}},
\end{equation}
where $e^{\lambda}_{ij}(\mathbf{k})$ is the polarization tensor corresponding to the mode with wave vector $\mathbf{k}$, and $\lambda$ denotes the two polarization states of the gravitational waves. The mode functions $h_{\mathbf{k}}^{\lambda}(\eta)$ satisfy the following inhomogeneous equation~\cite{Sorbo:2011rz, Caprini:2014mja, Sharma:2019jtb}
\begin{equation}\label{gweq}
    h_{\mathbf{k}}^{\lambda''} + 2\frac{a'}{a} h_{\mathbf{k}}^{\lambda'} + k^2 h_{\mathbf{k}}^{\lambda} = \mathcal{S}_{\mathbf{k}}^{\lambda},
\end{equation}
where primes denote differentiation with respect to conformal time $\eta$, and the source term $\mathcal{S}_{\mathbf{k}}^{\lambda}$ is given by~\cite{Maiti:2024nhv}
\begin{align}
    \mathcal{S}_{\mathbf{k}}^{\lambda}(\eta) = -\frac{2}{M_{\text{Pl}}^2} e^{ij}_{\lambda}(\mathbf{k}) \int \frac{\mathrm{d}^3 \mathbf{q}}{(2\pi)^{3/2}} \left[E_i(\mathbf{q}, \eta) E_j(\mathbf{k} - \mathbf{q}, \eta) + B_i(\mathbf{q}, \eta) B_j(\mathbf{k} - \mathbf{q}, \eta) \right].
\end{align}

The power spectrum of tensor fluctuations is defined as~\cite{Maiti:2024nhv}
\begin{align}\label{gen-def-pow}
    \langle h_{\mathbf{k}}^{\lambda} h_{\mathbf{k}'}^{\lambda'} \rangle = \frac{2\pi^3}{k^3} \mPt^{\lambda}(k, \eta) \delta^{(3)}(\mathbf{k} + \mathbf{k}').
\end{align}

The total tensor power spectrum consists of two independent contributions, one arising from vacuum fluctuations and another sourced by the electromagnetic field. These components are statistically independent and can be expressed as $\mPt=\mPtp+\mPts$,
where $\mPtp$ represents the tensor power spectrum from vacuum fluctuations, while $\mPts$ corresponds to the secondary tensor spectrum sourced by the electromagnetic field.

For simplicity, the source-induced tensor power spectrum can be expressed in terms of the spectral energy densities of the electric and magnetic fields, $\mPb(k,\eta)$ and $\mPe(k,\eta)$, as~\cite{Maiti:2025cbi}
\begin{align}
    \mPts(k,\eta) = \frac{2}{\Mp^4} \int_{0}^{\infty}\frac{\mathrm{d}q}{q}\int_{-1}^{1}\mathrm{d}\gamma \frac{f(\gamma,\beta)}{[1+(q/k)^2-2\gamma(q/k)]^{3/2}}
    \times \left[\int_{\eta_i}^{\eta} \mathrm{d}\eta_1 \ a^2(\eta_1) \mathcal{G}_k(\eta,\eta_1) \mPbe^{1/2}(q\eta_1) \mPbe^{1/2}(|\mathbf{k}-\mathbf{q}|\eta_1)\right],
\end{align}
where $f(\gamma,\beta) = (1+\gamma^2)(1+\beta^2)$, with $\gamma = \hat{\mathbf{k}} \cdot \hat{\mathbf{q}}$ and $\beta = \widehat{\mathbf{k}-\mathbf{q}} \cdot \hat{\mathbf{k}}$~\cite{Sharma:2019jtb}. Here, $\eta_i$ is the initial time when the source is activated, and $\eta$ is the time at which the tensor power spectrum is evaluated.

There are two primary stages where the production of the tensor power spectrum is significant, first, during reheating, when the coupling further amplifies the gauge field, and second, during the radiation-dominated era. Since the generation of the magnetic field reaches its peak at the end of reheating and ceases thereafter, both the magnetic field and the background evolve as $a^{-4}$ in this era. Due to the significant production of the magnetic field, we also find considerable generation of tensor perturbations during the radiation-dominated phase.

Before discussing the production of the tensor power spectrum sourced by electromagnetic fields, we provide a brief review of the primary generation of tensor perturbations (for a review, see~\cite{Starobinsky:1979ty, Grishchuk:1974ny, Guzzetti:2016mkm, Haque:2021dha, 1979ZhPmR..30..719S}).

\subsection{Production of Primary GWs}

It is well known that quantum fluctuations in spacetime, amplified during inflation, imprint a stochastic gravitational wave background (SGWB) that carries signatures of inflationary dynamics. Here, we consider a well-known slow-roll inflation model, where the universe follows a de Sitter expansion with the scale factor evolving as  $ a(\eta) = (1 - \HI \eta)^{-1}$, 
where \( \HI \) represents the nearly constant Hubble parameter during inflation. The homogeneous solution of Eq.~\eqref{gweq}, satisfying the Bunch-Davies initial condition, yields the tensor perturbation \( \hk \) (accounting only for vacuum fluctuations) as~\cite{Haque:2021dha}
\begin{align}
    \hk(\eta) = \frac{\sqrt{2}}{\Mp} \frac{i \HI}{\sqrt{2k^3}} \left[ 1 - \frac{i k}{\HI a(\eta)} \right] e^{-i k / \HI} e^{i k / (\HI a(\eta))}.\label{eq:hkp_inf}
\end{align}
Here, we observe that the amplitude of tensor perturbations is proportional to the inflationary energy scale, i.e., \( \hk \propto \HI \).  

After inflation, the universe undergoes different evolutionary phases. Depending on the background evolution, the initial fluctuations, which were amplified from the quantum vacuum, get modified. If we consider a non-instantaneous reheating scenario with an equation of state \( \wre \), the scale factor evolves as  $ a(\eta) =\are\left( {\eta}/{\ere} \right)^{\delta/2}$,  
where \( \delta(\wre) = {4}/{(1+3\wre)} \), and \( \are \) represents the scale factor at the end of inflation, i.e., at \( \eta = \eta_{\text{re}} \). In this background, with proper initial conditions, the solution of the tensor perturbation during reheating is given by~\cite{Maiti:2025cbi}
\begin{align}
    \hk(\eta > \ee) \simeq \frac{\pi^2 x^l}{2 \Gamma(l) \Gamma(1 - l)} \left[ \frac{2 - l}{\Gamma(2 - l)} \left( \frac{k}{k_{\text{re}}} \right)^{2(1 - l)} J_l(x) + \frac{1}{\Gamma(l)} J_{-l}(x) \right] \hk(\ee),
\end{align}
where we define \( x = k\eta \) and \( l(w_{\text{re}}) = {3(\wre - 1)}/{2(1+3w_{\text{re}})} \). Here, \( \hk(\ee) \) is the value of the tensor perturbation at the transition from inflation to reheating, governed by Eq.~\eqref{eq:hkp_inf}. The functions \( J_l(x) \) are Bessel functions of the first kind. From the above expression, we see that the evolution of tensor perturbations is affected by the background evolution during reheating.  

After reheating, the universe enters a radiation-dominated phase, where the background energy density scales as \( a^{-4} \). During this phase, the scale factor evolves as \( a(\eta) \propto \eta \), and the homogeneous solution of Eq.~\eqref{gweq} takes the form~\cite{Maiti:2025cbi}
\begin{align}
    \hk(\eta > \eta_{\text{re}}) = x^{-1} \left( \mathcal{D}_1 e^{-ix} + \mathcal{D}_2 e^{ix} \right),
\end{align}
where \( \mathcal{D}_1 \) and \( \mathcal{D}_2 \) are constants given by
\begin{subequations}\label{eq:D}
    \begin{align}
        \mathcal{D}_1(\xre) &= \frac{\hkr(\xre) \left( i \xre - 1 \right) - \xre \hkrp(\xre)}{2i} e^{i \xre}, \\
        \mathcal{D}_2(\xre) &= \frac{\xre \hkrp(\xre) + \hkr(\xre) \left( i \xre + 1 \right)}{2i} e^{-i \xre}.
    \end{align}
\end{subequations}
Using Eq.~(\ref{eq:D}), we define the tensor power spectrum during the radiation-dominated era (accounting only for homogeneous contributions) as~\cite{Maiti:2025cbi}
\begin{align}
    \mathcal{P}(k, \eta > \eta_{\text{re}}) \simeq \frac{1}{2 (k \eta)^2} \left( 1 + \left( \frac{k}{k_{\text{re}}} \right)^2 \right) \mathcal{P}(k, \eta_{\text{re}}).
\end{align}
Here, \( \mathcal{P}(k, \ere) \) represents the total tensor power spectrum at the end of reheating. For modes that remain outside the horizon at the end of reheating (\( k < k_{\text{re}} \)), the tensor power spectrum in the radiation-dominated era scales proportionally to the inflationary tensor power spectrum, i.e.,  $\mathcal{P}(k, \eta > \ere) \propto \mathcal{P}(k, \ere)$.
However, for modes that re-entered the horizon before the end of reheating (\( k > \kre \)), the tensor power spectrum is influenced by the reheating history, scaling as $ \mathcal{P}(k > \kre, \eta > \ere) \propto \xre^{2l + 1} \mathcal{P}(k, \ere),$ where \( \xre = k \ere \)~\cite{Maiti:2025cbi}.

\subsection{Production of Secondary GWs}
In this subsection, we discuss the production mechanisms of secondary gravitational waves (SGW) sourced by the electromagnetic (EM) field generated during and after inflation. As analyzed in the earlier sections, for this type of coupling, where both the backreaction issue and the strong coupling problem are addressed simultaneously, the model parameters must be chosen such that the large-scale magnetic field generated during inflation remains subdominant compared to the background energy density.  

However, since the coupling function remains active during reheating, we find a significant enhancement in both the magnetic and electric energy densities, even for modes that remain outside the horizon before the end of reheating. Notably, the production of the magnetic field reaches its maximum at the end of reheating, where the coupling function satisfies \( f(\eta = \eta_{\text{re}}) = 1 \). This allows us to neglect the contribution of tensor perturbations sourced by the EM field during inflation.  

Thus, the dominant production of secondary GWs occurs in two distinct phases:  
\begin{itemize}
    \item \textbf{During reheating}, when the amplified EM energy density acts as a source for tensor perturbations.  
    \item \textbf{During the radiation-dominated era}, where any remaining sourced fields (magnetic field) continue to contribute to GW production.
\end{itemize}

\paragraph{\underline{Generation of Tensor Perturbations During Reheating}}\

For super-horizon modes, the electric field energy density dominates over the magnetic field. Once the modes re-enter the horizon, both the electric and magnetic fields exhibit the same spectral behavior with equal amplitude (see Fig.~\ref{fig:Pbe_vs_k_nb}). Since we are computing the tensor power spectrum generated during reheating, we can simplify our analysis by neglecting the magnetic field contribution for now. 

The tensor power spectrum induced by the electric field during reheating, evaluated at $\eta = \ere$, is given by
\begin{align}
    \mPts(k,\eta_{\rm re}) = \frac{2}{\Mp^4 k^4} \int_{u_{\rm min}}^{u_{\rm max}} \frac{du}{u} \int_{-1}^{1} d\mu \frac{f(\mu, \gamma)}{[1+u^2-2\mu u]^{3/2}} \left[ \int_{\xe}^{\xre} dx_1 a^2(x_1)\mGk(\xre, x_1) \mPe^{1/2}(ux_1) \mPe^{1/2}(|1-u|x_1) \right]^2
\end{align}

Here, we define the dimensionless variable $u = q/k$ and $\xre = k\ere$, where $\ere$ is the conformal time at the end of reheating. The above integral does not admit an exact analytical solution, but we can rewrite it in the following approximate form
\begin{align}
    \mPts(k,\ere) \simeq \frac{|\mB|^4}{32} \left(\frac{\HI}{\Mp}\right)^4 \left(\frac{k}{\ke}\right)^{2(\delta - 2n - 2\alpha)} \left\{ \overline{\mathcal{I}_1(k,\ere}) + \overline{\mathcal{I}_2(k,\ere)} \right\},
\end{align}
where the terms $\mathcal{I}_1$ and $\mathcal{I}_2$ are defined as
\begin{subequations}
    \begin{align}
        \mathcal{I}_1(k,\ere) &= \int_{u_{\rm min}}^1 du\, u^2 u^{-2(n+\alpha)} \int_{-1}^1 d\gamma\, f(\gamma, \beta) \left[ \int_{\xe}^{\xre} dx_1 x_1^{1-\delta} \mGk(\xre, x_1) J_{\alpha - \frac{1}{2}}(ux_1) J_{\alpha - \frac{1}{2}}(x_1) \right]^2, \\
        \mathcal{I}_2(k,\ere) &= \int_1^{u_{\rm max}} du\, u^2 u^{-4(n+\alpha)} \int_{-1}^1 d\gamma\, f(\gamma, \beta) \left[ \int_{\xe}^{\xre} dx_1 x_1^{-1-\delta} \mGk(\xre, x_1) J_{\alpha - \frac{1}{2}}^2(ux_1) \right]^2.
    \end{align}
\end{subequations}

In these expressions, we have defined $u_{\rm max} = \ke/k$ and $u_{\rm min} = \kpv/k$, where $\kpv$ is the pivot scale observed the CMB, and $\ke$ is the highest mode leaving the horizon at the end of inflation. The overline indicates the oscillation average of the corresponding quantity.

Since an exact analytical solution is not feasible, we evaluate the spectral behavior by considering specific limits. We are particularly interested in large-scale magnetic fields that could explain the presently observed cosmic magnetic field. For super-horizon scales ($k \ll \kre$), we estimate the tensor power spectrum as
\begin{align}
 \mPts(k \ll \kre, \ere) \simeq \frac{|\mB|^4}{48} \frac{2^{2-4\alpha} \Gamma^2(l)}{\Gamma^4(\alpha+\frac{1}{2}) \Gamma^2(1+l)} \left(\frac{\HI}{\Mp}\right)^4 \frac{1 - \left(\frac{\kpv}{k}\right)^{2-2n}}{(2-2n)(2\alpha - \delta + 2)} \left(\frac{\xe}{\xre}\right)^{2(\delta - 2\alpha)} \left(\frac{k}{\ke}\right)^{4(1-n)}.
 \label{eq:ptre_super}
\end{align}

From Eq.~\eqref{eq:ptre_super}, we observe that the tensor power spectrum due to the electric field follows $\mPts(k \ll \kre, \ere) \propto k^{4(1-n)}$, which is twice the spectral index of the electric field, i.e., $\mPe(k) \propto k^{2(1-n)}$ (see Eq.~\eqref{eq:pe_super}). This result implies that for $n > 1$, the tensor power spectrum exhibits a red-tilted spectrum, leading to an overproduction of tensor fluctuations at large wavelengths. Consequently, this could violate the current observational bound on the tensor-to-scalar ratio, $r_{0.05} \leq 0.036$~\cite{Planck:2015fie, BICEP:2021xfz, BICEP2:2018kqh, Planck:2018jri}.

For instance, considering $\wre = 0$ and a reheating temperature of $\Tre = 1$ GeV, we find that to satisfy the observational bound on $r$, the coupling parameter must be constrained as $n_{\rm max} \leq 1.07$. However, for the same set of parameters, we find that the produced electromagnetic fields can easily backreact with the background (see Fig.~\ref{fig:delta_n}).

\paragraph{\underline{Generation of tensor perturbation during radiation-dominated era:}}\

After reheating, there is no further production of the gauge field, but during the reheating era, the production of the gauge field is significant. During reheating, all fundamental particles are produced from the decay of the inflaton field, and the temperature of the universe was very high. Consequently, we can assume that the conductivity during the radiation-dominated era was also very high. In the presence of high conductivity, the electric field decays quickly due to the rapid response of charged particles (mainly electrons) in the plasma medium, while the magnetic field freezes (if we ignore magnetohydrodynamics). As we have seen, the magnetic field strength at the end of reheating is quite high at small scales near $k\simeq\kre$ (see Fig.~\ref{fig:pbe_vs_k_tre}), and it can further produce tensor fluctuations. However, the source due to the magnetic field is only effective up to the neutrino decoupling era, i.e., $\eta=\eta_\nu$, because after neutrinos decouple from the thermal bath, they can balance the anisotropies induced by the magnetic field \cite{PhysRevD.70.043011, KAGRA:2021kbb}. Therefore, for the radiation-dominated era, the time range should be from $\xre$ to $x_\nu=k\eta_\nu$.
\begin{align}
\mPts(k,\eta_{\nu})= \frac{2}{\mpl^4 k^4}\left( \int_{\xre}^{x_\nu} dx_1 \frac{\mathcal{G}{\rm ra}(x_\nu,x_1)}{a^2(x_1)}\right)^2 \times \int_0^{\infty}\frac{dq}{q}\int_{-1}^1 d\mu f(\mu,\beta) \frac{\tilde{\pb}(k,\ere)\tilde{\pb}(|\vk-\vq|\ere)}{[1+(q/k)^2-2\mu(q/k)]^{3/2}} \label{eq:pt_ra}
\end{align}
Here, the tilde denotes the comoving quantity, i.e., $\tilde{\mPb}(k,\ere)=a^4(\eta)\mPb(k,\eta)$. This is convenient because, after reheating, there is no further production of the gauge field. Due to high conductivity and the absence of an electric field, the magnetic field evolves adiabatically and, due to background expansion, dilutes as $a^{-4}$. To facilitate the time integral separately, we express the transient power spectrum in terms of these comoving quantities.

Utilizing Eq.~\eqref{eq:pb_sub_h}, we can define the comoving magnetic spectral energy density $\tilde{\mPb}$ as
\begin{align}
\tilde{\mPb}(k,\eta>\ere)=\frac{k^4|\mB|^2}{8\pi}\left(\frac{k}{\ke}\right)^{-2(n+\alpha+1)}\left(\frac{k}{\kre}\right)\bJ^2_{\alpha+\frac{1}{2}}\left(k/\kre\right)\label{eq:c_pbe}
\end{align}
here we identify $\mB$ as
\begin{align}\label{eq:def_beta}
    \mB=\frac{2^{n+\alpha}(\alpha-n)\Gamma\l(n+\frac{1}{2}\r)}{\cos(\pi\alpha)\Gamma\l(\frac{1}{2}-\alpha\r)}
\end{align}

Now, we can write the tensor power spectrum produced during the radiation-dominated era as
\begin{align}
\mPtsra(k,\eta) &= \frac{k^4|\mB|^4}{32\pi^2\Mp^4}\left( \frac{k}{\ke} \right)^{-4(n+\alpha+1)}\xre^2 \left( \int_{\xre}^{x_{\nu}} dx_1 \frac{\Gk(x_\nu,x_1)}{a^2(x_1)}\right)^2\nn\\
&\times \int_{u_{\rm min}}^{u_{\rm max}} \frac{du}{u}\int_{-1}^1 d\mu \frac{f(\mu,\gamma)}{[1+u^2-2u\mu]^{3/2}}u^{3-2(n+\alpha)}(|1-u|)^{3-2(n+\alpha)}\mJ^2_{\alpha+1/2}(u\xre)\mJ^2_{\alpha+1/2}(|1-u|\xre)
\end{align}
We can numerically solve the above integral by performing the time and momentum integrals separately. However, to analyze the spectral behavior of the tensor power spectrum, we can also evaluate it analytically by considering two different limits: (i) the spectral behavior of the tensor power spectrum for modes that remain outside the horizon before the end of reheating ($k<\kre$), and (ii) the spectral behavior of modes that are inside the horizon before the end of reheating ($k>\kre$).

\paragraph{\underline{Computing the Tensor Power Spectrum for $k<\kre$}:}
For modes that remain outside the horizon before the end of reheating ($k<\kre$), we can take the limit $\xre=k/\kre<1$ to simplify the integral. We also split the momentum integral into two parts, $\kcmb/k<u<1$ and $1<u<\ke/k$. Utilizing this, we can express the tensor power spectrum as
\begin{align}
\mPtsra(k,\eta_\nu)= \frac{k^4|\mB|^4}{32\pi^2\Mp^4}\left( \frac{k}{\ke} \right)^{-4(n+\alpha+1)}\xre^2 \left( \int_{\xre}^{x_{\nu}} dx_1 \frac{\Gk(x_\nu,x_1)}{a^2(x_1)}\right)^2 \times \left\{ F_{uu}^1(k,\ere)+F_{uu}^2(k,\ere)\right\}
\end{align}
where we define
\begin{align}
F_{uu}^1(k,\ere)&=  \int_{u_{\rm min}}^1 \frac{du}{u} \int_{-1}^1 d\mu f(\mu,\gamma) u^{3-2(n+\alpha)}\mJ^2_{\alpha+1/2}(u\xre)\mJ^2_{\alpha+1/2}(\xre)\label{eq:fuu1} \\
F^2_{uu}(k,\ere)&= \int_1^{u_{\rm max}}\frac{du}{u^4}\int_{-1}^1d\mu f(\mu,\gamma)u^{6-4(n+\alpha)}\mJ^4_{\alpha+1/2}(u\xre)\label{eq:fuu2}
\end{align}
Numerical evaluation shows that $F_{uu}^1(k,\ere) \gg F_{uu}^2(k,\ere)$ (see Appendix~\ref{ape_1}). Focusing only on $F_{uu}^1$, in the limit $\xre\ll1$ with $\kcmb/k<u<1$, we find
\begin{align}
F^1_{uu}(k,\ere)\simeq \frac{8}{3}\frac{2^{-4(\alpha+1/2)}}{\Gamma^4(\alpha+1/2)}\xre^{4(\alpha+3/2)} \frac{1}{4-2n} \left\{ 1 -\left( \frac{k_0}{k} \right)^{4-2n}\right\}
\end{align}

\subsection{Defining the Present-Day Dimensionless Gravitational Wave Spectrum Energy Density $\ogwh$}

Gravitational waves (GWs) weakly interact with matter and can freely propagate without losing their original nature after being produced during a radiation-dominated era. The GW energy density, which scales as $\rho_{\text{GW}} \propto a^{-4}$, can be normalized by the total energy density at the production time, $\rho_c(\eta)$. Presently, this density parameter is given by
\begin{align}
    \ogw(k, \eta) = \frac{\rho_{\text{GW}}(k, \eta)}{\rho_c(\eta)} = \frac{1}{12} \frac{k^2 \Pt(k, \eta)}{a^2(\eta) H^2(\eta)},
\end{align}
where $\rho_c(\eta) = 3 H^2(\eta) \Mp^2$, with the reduced Planck mass $\Mp \approx 2.43 \times 10^{18} \, \Gev$.

The GW energy density follows the radiation scaling behavior, making modes within the Hubble radius near radiation-matter equality particularly relevant. The present-day energy density parameter $\ogw(k) h^2$ is then given by
\begin{align}
    \ogw(k) h^2 \simeq \left(\frac{\gsp}{\gseq}\right)^{1/3} \Omega_R h^2 \ogw(k, \eta),
\end{align}
where $\Omega_R h^2 = 4.3 \times 10^{-5}$, $\gseq \simeq \gsp = 3.35$, and denote relativistic degrees of freedom at equality and today, respectively.

The spectral energy density (SED) of GWs helps us distinguish between different production mechanisms, such as vacuum fluctuations and sourced fields like the electromagnetic field.

\paragraph{\underline{Spectral Behavior of Primary GWs:}}
Primordial gravitational waves (PGWs) originate from vacuum fluctuations during the inflationary era. As previously discussed, these tensor perturbations exhibit a nearly scale-invariant spectrum in a de Sitter inflationary background. However, the present-day amplitude and shape of the PGW spectrum are influenced by the post-inflationary evolution of the universe. In particular, non-standard reheating scenarios can significantly modify the spectral behavior of GWs, especially for modes that re-enter the horizon before the end of reheating. Consequently, the present-day spectrum of PGWs, depending on the comoving wavenumber, can be characterized by the following expression~\cite{Maiti:2025cbi, Maiti:2024nhv, Chakraborty:2024rgl}
\begin{align}
\ogwp(k) h^2\simeq  
\frac{\Omega_{\rm r}h^2\, \gsp^{1/3}}{6\,\gseq^{1/3}}\frac{\HI^2}{\Mp^2}  
\times\left\{\begin{array}{ll} 1 & k<\kre,\\
\mathcal{D}_1\l(\frac{k}{\kre}\r)^{-\nw} & k>\kre,
\end{array}\right.
\end{align}
where $\mD\simeq 2^{1-2l}\Gamma^2(1-l)/2\pi\simeq \mathcal{O}(1)$ and $\nw=2(1-3\wre)/(1+3\wre)$~\cite{Maiti:2025cbi, Maiti:2024nhv}.

For GWs produced solely from vacuum fluctuations, modes remaining outside the horizon before reheating end are not affected by reheating dynamics and retain a scale-invariant spectrum regardless of the equation of state (EoS). However, for modes that re-enter the horizon before reheating ends, the spectrum depends on the reheating phase dynamics. Specifically, for $\wre=0$ (matter-like reheating), the spectrum follows $k^{-2}$ for $k>\kre$, whereas for $\wre=1/3$, it remains scale-invariant. For $\wre>1/3$, the spectrum transitions to a blue-tilted behavior, scaling as $k^{-\nw}$. For instance, with $\wre=0.5$, the spectrum scales as $k^{0.4}$.

\paragraph{\underline{Spectral Behavior of Secondary GWs:}}
Similarly, for GWs sourced by electromagnetic fields, we approximate the SED of secondary GWs as
\begin{align}
 \ogw^{\rm sec}h^2 &\simeq \frac{\Omega_{\rm r}h^2}{6}\l(\frac{\gsp}{\gseq}\r)^{1/3}\l(\frac{\HI}{\Mp}\r)^4\l(\frac{\ae}{\are}\r)^{2(1-3\wre)}\l(\frac{\xre}{\xe}\r)^{4(\alpha+1)}\overline{\mI^2(k,\ere,\eta_\nu})\nn\\
& \times \l\{
    \begin{matrix}
       \mathcal{A}_1 (k/\ke)^{2(4-2n)} & \kcmb<k<\kre\\
       \mathcal{A}_2 (\xre/\xe)^{4(n-2)}(k/\ke)^{4(1-n-\alpha)} & \kre<k<\ke
    \end{matrix}
    \r.
\end{align}
where we define $\mathcal{A}_1$ and $\mathcal{A}_2$ as
\begin{subequations}
    \begin{align}
        \mathcal{A}_1 &=\frac{|\mB|^42^{-4(n+1/2)}}{144\,\pi^2(4-2n)\Gamma^4(\alpha+1/2)},\\
        \mathcal{A}_2 &=\frac{|\mB|^4}{360\,\pi^4}\frac{1}{2(n+\alpha)-2},
    \end{align}
\end{subequations}
where $\mB$ is defined in Eq.~\eqref{eq:def_beta}. Here, $\mIra(k,\kre,\knu)$ is defined as~\cite{Maiti:2024nhv}
\begin{align}
\mI(k,\ere,\eta_\nu)=\int_{\xre}^{x_\nu}dx_1\frac{\sin(x_1 - x_\nu)}{x_1} .
\end{align}
As for the low-frequency regions, we found that $\mI(k<\knu)=\gamma_1$, where $\gamma_1\sim 0.5$ at the epoch of neutrino decoupling.

In the above expression, the factor $\left({\ae}/{\are}\right)^{2(1-3\wre)}$ arises from the relative dilution of the magnetic field energy density and background energy density. The magnetic energy density scales as $a^{-4}$, whereas the background energy density scales as $a^{-3(1+\wre)}$. The factor $\left({\xre}/{\xe}\right)^{4(\alpha+1)}$ accounts for additional magnetic field production due to coupling during reheating.

For super-horizon modes, the spectral shape of secondary GWs is governed by the initial gauge field coupling during inflation via $n$, yielding a spectral index twice the magnetic spectral index for super-horizon modes (see Eq.~\eqref{eq:pb_sup_h}). For sub-horizon modes, the spectral tilt is also twice the magnetic spectral index, scaling as $4(1-n-\alpha)$ (see Eq.~\eqref{eq:pb_sub_h}). This follows naturally since the SED of secondary GWs is proportional to the square of the magnetic spectral energy density (see Eq.~\eqref{eq:pt_ra}), i.e., $\ogw(k,\eta)\propto \mPb^2(k,\eta)$.

 \begin{figure*}
\includegraphics[width=0.45\linewidth]{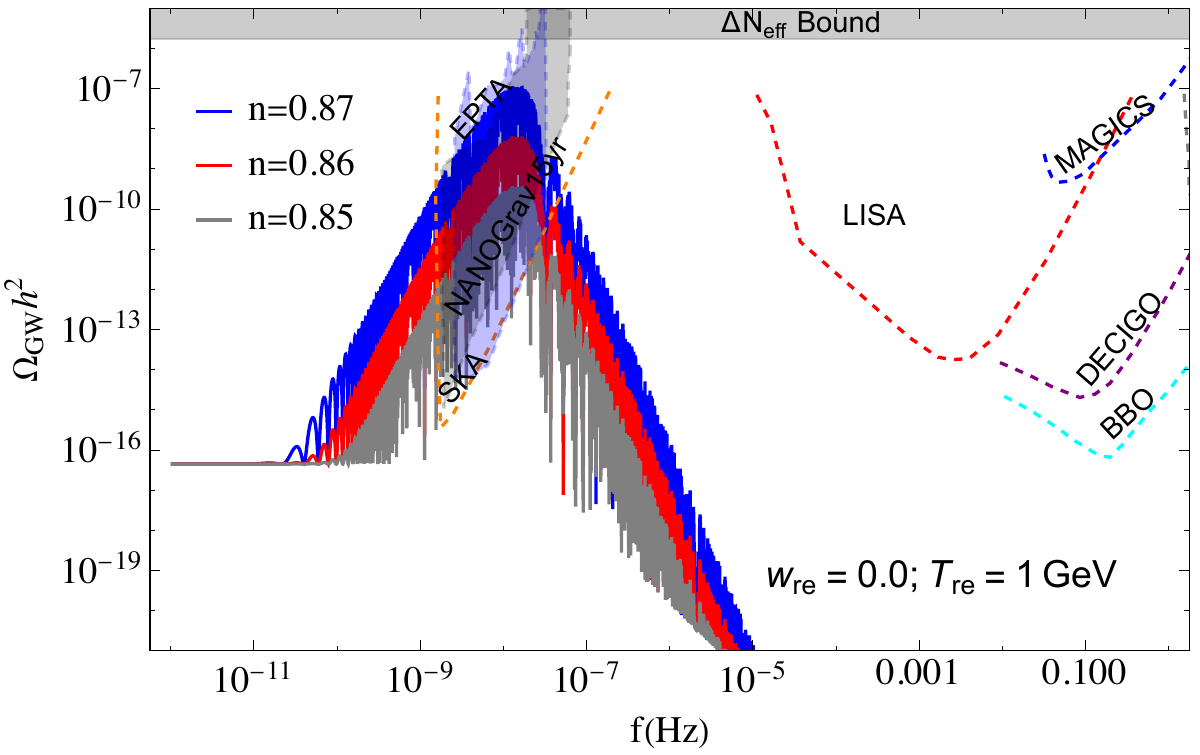}
\includegraphics[width=0.45\linewidth]{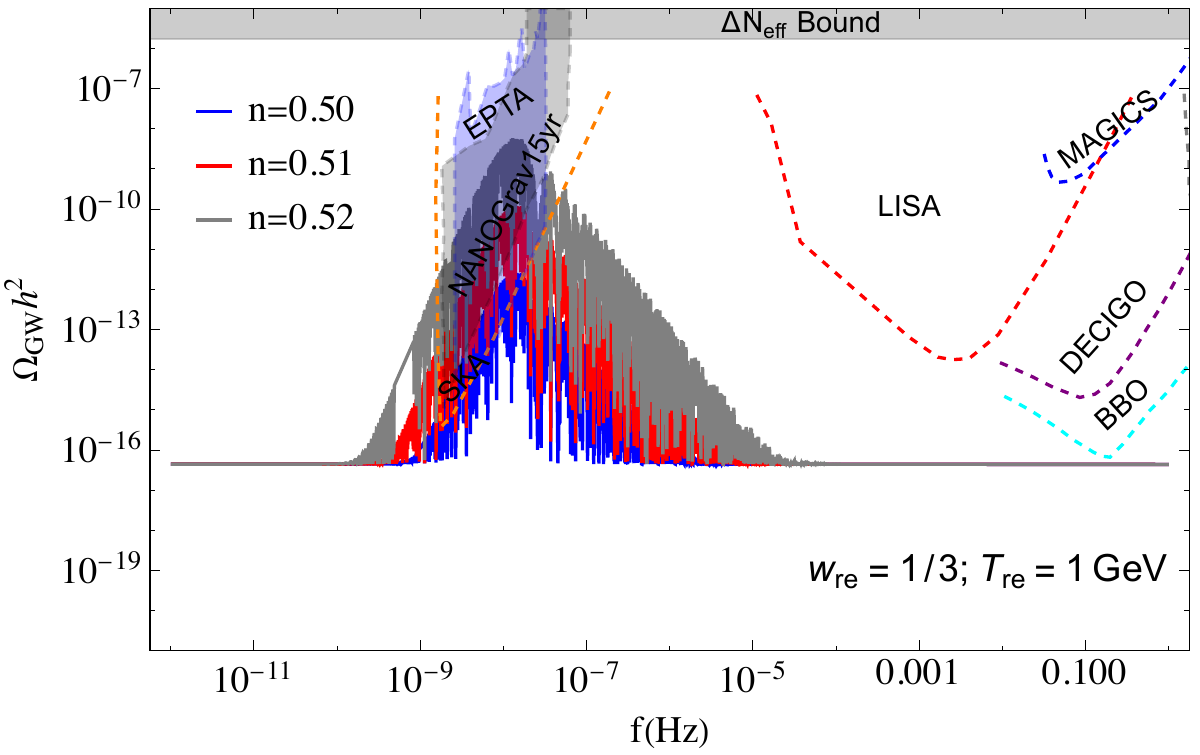}
\caption{Gravitational wave spectral energy density \( \ogwh \) as a function of frequency \( f \) (in Hz) for different scenarios. In the left panel, we consider a fixed reheating scenario with \( \Tre = 1 \) GeV and \( \wre = 0 \), varying the coupling parameter \( n \) (\( n = 0.85, 0.84, 0.83 \)). In the right panel, we fix the coupling parameter at \( n = 0.78 \) and the equation of state at \( \wre = 0 \), while varying the reheating temperature \( \Tre \) (\( \Tre = 100, 50, 10 \) GeV). Different colors represent different parameter values in both figures.} 
\label{fig:gw1}
\end{figure*}
 \begin{figure*}
\includegraphics[width=0.45\linewidth]{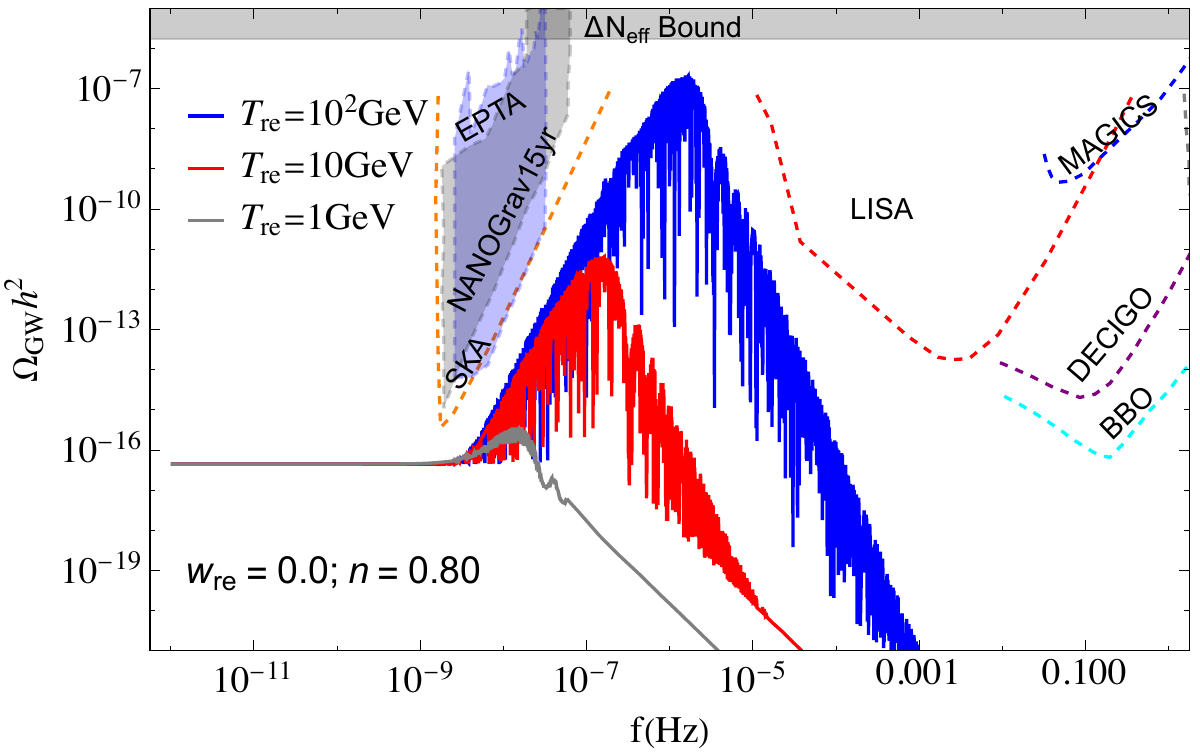}
\includegraphics[width=0.45\linewidth]{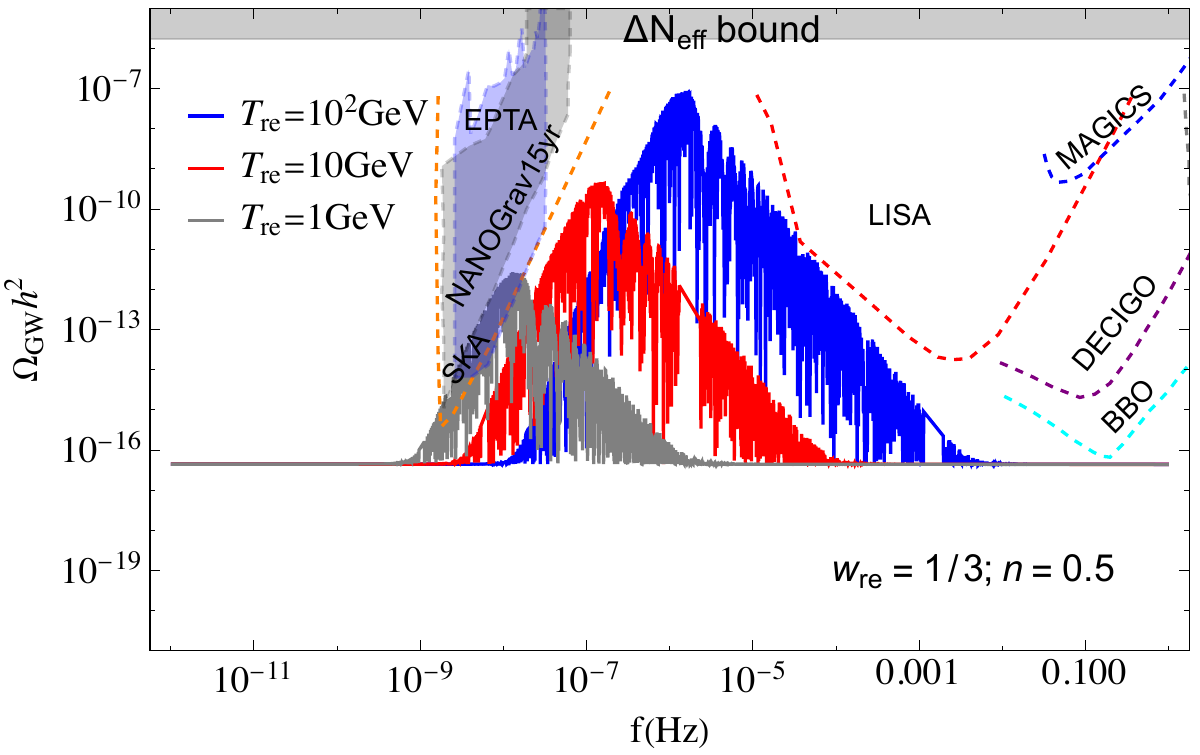}
\caption{Gravitational wave spectral energy density \( \ogwh \) as a function of frequency \( f \) (in Hz) for different scenarios. In the left panel, we consider a fixed reheating scenario with \( \Tre = 1 \) GeV and \( \wre = 0 \), varying the coupling parameter \( n \) (\( n = 0.85, 0.84, 0.83 \)). In the right panel, we fix the coupling parameter at \( n = 0.78 \) and the equation of state at \( \wre = 0 \), while varying the reheating temperature \( \Tre \) (\( \Tre = 100, 50, 10 \) GeV). Different colors represent different parameter values in both figures.} 
\label{fig:gw2}
\end{figure*}

In Fig.~\ref{fig:gw1}, we present the spectral energy density (SED) of gravitational waves (GWs), $\Omega_{\rm GW}$, as a function of the comoving frequency $f$ (Hz) for two different values of the equation of state (EoS) during reheating, $\wre = 0.0$ (left) and $\wre = 1/3$ (right). In both cases, we set the reheating temperature to $\Tre=1$ GeV. This choice ensures consistency with the observed present-day magnetic field at large scales, particularly at $1$ Mpc. 

From Table~\ref{tab:param_est}, it is evident that for this class of magnetogenesis models to remain free from backreaction and strong coupling issues, the reheating temperature must be relatively low. For the left panel ($\wre = 0$), we find that to obtain a reasonable magnetic field strength at $1$ Mpc, the coupling parameter must lie within the range $0.8 < n < 0.95$, with the reheating temperature constrained to $10^{-2} \leq \Tre\leq 10^2$ GeV. Within this parameter range, the generated magnetic fields source a significant secondary GW background, detectable not only by future SKA~\cite{Janssen:2014dka} but also by current PTA observations, including NANOGrav~\cite{NANOGrav:2023gor}, EPTA+IPTA~\cite{2023arXiv230616224A}, PPTA~\cite{Reardon:2023gzh, Zic:2023gta}, and CPTA~\cite{Xu:2023wog}.

Although the latest NANOGrav 15-year data suggests a characteristic strain spectral index $\gamma_{\rm CP} = 3.2$ (corresponding to an energy density spectral index $n_{\rm gw} = 1.8$), our magnetogenesis model naturally predicts a strongly blue-tilted magnetic field spectrum, scaling as $\mathcal{P}_B(f) \propto f^{4-2n}$. For instance, choosing $n = 0.85$ gives $\mathcal{P}_B(f) \propto f^{2.3}$. Since for super-horizon modes the secondary GW spectrum follows $\Omega_{\rm GW}(f) \propto f^{2(4-2n)}$, we obtain $\Omega_{\rm GW}(f) \propto f^{4.6}$ at nano-Hz frequencies for $T_{\rm re} = 1$ GeV, indicating a highly blue-tilted GW spectrum.

In the right panel of Fig.~\ref{fig:gw1}, we consider $w_{\rm re} = 1/3$ with $T_{\rm re} = 1$ GeV. To avoid backreaction effects, we select $n = 0.50, 0.51, 0.52$. These values yield sufficient magnetic fields for detection by future experiments such as SKA. In this case, for super-horizon modes ($k < k_{\rm re}$), the GW spectrum scales as $\Omega_{\rm GW}(f) \propto f^6$ for $n = 0.5$. 

Interestingly, we always observe a peak in the GW spectrum around $f \simeq \fre $, corresponding to the maximum amplitude of the electromagnetic field at $k \simeq \kre$. Identifying this peak frequency could provide insights into the background temperature at the end of reheating. Additionally, we find that the spectral slope of the sub-horizon modes is influenced not only by the coupling parameter $n$ but also by the details of the reheating dynamics. Specifically, for sub-horizon modes, the GW spectrum follows $\ogwh(f) \propto f^{4(1-n-\alpha)}$, where $\alpha=2n\NI/(1+3\wre)\Nre$ depends on the equation of state and reheating temperature. Our analysis indicates that when the reheating dynamics resemble a radiation-dominated background, the resulting magnetic spectrum is more strongly blue-tilted, leading to a correspondingly steeper GW spectrum compared to matter-like reheating scenarios.

In Fig.~\ref{fig:gw2}, we further explore the impact of reheating temperature by plotting the SED of GWs for $w_{\rm re} = 0.0$ (left) and $w_{\rm re} = 1/3$ (right) for three different reheating temperatures: $T_{\rm re} = 10^2$ GeV (blue), $T_{\rm re} = 10$ GeV (red), and $T_{\rm re} = 1$ GeV (gray). For $w_{\rm re} = 0.0$, we fix $n = 0.80$, while for $w_{\rm re} = 1/3$, we set $n = 0.5$. 

Our results show that both the amplitude and peak position of the GW spectrum vary significantly with reheating temperature. A higher reheating temperature implies an earlier end to reheating. To satisfy the condition $f(\eta =\ere) = 1$, a larger $\Tre$ requires a higher value of $\alpha$, which enhances electromagnetic field production during reheating. Consequently, a higher reheating temperature leads to a stronger peak in the magnetic field spectrum (see Fig.~\ref{fig:delta_tre}), thereby amplifying the secondary GW background. This trend is clearly illustrated in Fig.~\ref{fig:gw2}, where increasing $\Tre$ results in a larger GW amplitude and a shift in the spectral peak.

\section{Conclusion}\label{sec:conclusion}

The origin of large-scale magnetic fields, with wavelengths exceeding those of galaxies and galaxy clusters, remains an open question in cosmology. While several models have been proposed to explain their genesis, their consistency with observational constraints is still a subject of investigation. In this work, we explored a coupling mechanism that can simultaneously address the issue of strong coupling while avoiding backreaction effects, thereby providing a viable explanation for the observed present-day magnetic field strength.

Most studies on this topic assume an arbitrary post-inflationary universe; however, in realistic reheating scenarios, inflation and reheating are strongly correlated, limiting the range of permissible parameter values. Our analysis indicates that achieving the required present-day magnetic field strength is not possible by arbitrarily choosing model parameters. Even when obtaining the desired field strength, many models necessitate an unrealistically low inflationary energy scale. In contrast, we have demonstrated that by carefully selecting coupling parameters and considering specific reheating scenarios, one can successfully generate the observed large-scale magnetic fields without requiring an extremely low inflationary scale.

Many magnetogenesis models suggest that producing sufficiently strong magnetic fields on cosmological scales requires a scale-invariant or nearly scale-invariant spectrum. However, our findings indicate that even a strongly blue-tilted spectral behavior can still account for the observed magnetic field strength. A key requirement for this scenario is a reheating phase characterized by a low reheating temperature. Specifically, for a matter-like reheating scenario ($w_{\rm re} = 0$), we find that in order to satisfy the lower bound on the present-day magnetic field strength for wavelengths larger than $1\,\text{Mpc}$, the reheating temperature must lie within the range $10^{-2}T_{\rm re} < 10^2~\text{GeV}$, while the coupling parameter $n$ should be within the interval $0.8 < n < 0.95$. Similarly, for a radiation-like reheating scenario ($w_{\rm re} = 1/3$), we find that to generate a sufficiently strong magnetic field on large scales, the coupling parameter must satisfy $0.5 < n < 0.55$, with the reheating temperature constrained to $10^{-2} < T_{\rm re} < 10^2\,\text{GeV}$.

An intriguing outcome of our analysis is that the resulting magnetic field spectrum always exhibits a broken power-law behavior. We further observe that for matter-like reheating scenarios, the produced field, while still consistent with present-day observations, exhibits a smaller blue tilt compared to scenarios with a radiation-like reheating phase ($\wre = 1/3$). In both cases, however, our results consistently point toward a low reheating temperature as a key requirement for successful magnetogenesis.

In the second part of this study, we explored a potential observational test for this class of magnetogenesis models via the production of secondary gravitational waves (GWs). Our results show that for suitable model parameters, the generated electromagnetic (EM) field can serve as a significant source of GWs with a distinctive spectral signature. If the present-day magnetic field strength is to be satisfied, the parameter choices lead to a GW spectrum that peaks in the nano-Hz frequency range, making it compatible with the recently observed PTA signals \cite{NANOGrav:2023gor, 2023arXiv230616224A, Reardon:2023gzh, Zic:2023gta, Xu:2023wog}. Due to the strongly blue-tilted nature of the magnetic spectrum, the resulting GWs also inherit this characteristic, which can help stabilize the PTA signal within a $2\sigma$ limit.

Furthermore, even if we do not aim to explain the present-day large-scale magnetic field strength, our model allows for the generation of strong magnetic fields at smaller scales, which can act as prominent GW sources. Depending on the parameter choices, the resulting GW spectrum could be detectable by future sensitivity curves of LISA~\cite{Amaro-Seoane:2012aqc, Barausse:2020rsu}, DECIGO \cite{Seto:2001qf, Kawamura:2011zz, Suemasa:2017ppd}, or BBO \cite{Crowder:2005nr, Corbin:2005ny, Baker:2019pnp}. Notably, we find that generating such strong GWs does not require a low inflationary scale; signals consistent with $\HI\simeq 10^{-5} \Mp$ can still be easily achieved.

The primary objective of this work has been to investigate the impact of reheating dynamics on magnetogenesis models and their implications for the present-day magnetic field strength. Additionally, we have examined how the GW spectrum is influenced by these magnetogenesis scenarios and their associated reheating history. Our results indicate a distinct spectral behavior for the resulting GWs, which varies with the reheating dynamics, allowing for a clear differentiation between different scenarios. Moreover, the strongly blue-tilted GW spectrum predicted by these models makes them easily distinguishable from other sources, providing a potential observational signature for testing magnetogenesis models in the early universe.

 \acknowledgments
SM gratefully acknowledges financial support from the Council of Scientific and Industrial Research (CSIR), Ministry of Science and Technology, Government of India. SM also thanks Debaprasad Maity for valuable discussions and insightful comments.

\appendix
\subsection{\underline{Computing the tensor power spectrum for $k<\kre$:}}\label{ape_1}

Let us now consider those modes that remain outside the horizon at the end of reheating, i.e., \( k < \kre \). For such modes, the time integral contributing to the gravitational wave amplitude can be split into two domains: \( \umin \leq u \leq 1 \) and \( 1 \leq u \leq \umax \). In this regime, we can safely take the limit \( \xre < 1 \), as the scale factor at the end of reheating is still much smaller than today. This allows us to simplify the integral expression accordingly as:
\begin{align}
     \mPts(k,\eta_\nu)= &\frac{k^4|\mB|^4}{32\pi^2\Mp^4}\l( \f k \ke \r)^{-4(n+\alpha+1)}\xre^2 \l( \int_{\xre}^{x_{\nu}} dx_1 \frac{\Gk(x_\nu,x_1)}{a^2(x_1)}\r)^2\nn\\
     &\times \l\{ 
     \int_{u_{\rm min}}^1 \f {du} u \int_{-1}^1 d\mu \frac{f(\mu,\gamma)}{[1+u^2-2u\mu]^{3/2}} u^{3-2(n+\alpha)}\mJ^2_{\alpha+1/2}(u\xre)\mJ^2_{\alpha+1/2}(\xre)\r.\nn\\
     &\l. + \int_1^{u_{\rm max}}\frac{du}{u}\int_{-1}^1d\mu \frac{f(\mu,\gamma)}{[1+u^2-2u\mu]^{3/2}}u^{6-4(n+\alpha)}\mJ^4_{\alpha+1/2}(u\xre) \r\}
\end{align}
We can further simplify the above integral by separately considering the limits \( u < 1 \) and \( u > 1 \). This division corresponds to the convolution contributions from sub-horizon and super-horizon modes during reheating, enabling us to evaluate each portion of the integral with appropriate approximations valid in these respective domains.
\begin{align}
   \mPts(k,\eta_\nu)= \frac{k^4|\mB|^4}{32\pi^2\Mp^4}\l( \f k \ke \r)^{-4(n+\alpha+1)}\xre^2 & \l( \int_{\xre}^{x_{\nu}} dx_1 \frac{\Gk(x_\nu,x_1)}{a^2(x_1)}\r)^2\nn\\
    &\times \l\{ 
     \int_{u_{\rm min}}^1 \f {du} u \int_{-1}^1 d\mu f(\mu,\gamma) u^{3-2(n+\alpha)}\mJ^2_{\alpha+1/2}(u\xre)\mJ^2_{\alpha+1/2}(\xre) \r.\nn\\
     &\l. + \int_1^{u_{\rm max}}\frac{du}{u^4}\int_{-1}^1d\mu f(\mu,\gamma)u^{6-4(n+\alpha)}\mJ^4_{\alpha+1/2}(u\xre) \r\} \label{eq:ptra_i}
\end{align}
\begin{align}
    \mPts(k,\eta_\nu)= \frac{k^4|\mB|^4}{32\pi^2\Mp^4}\l( \f k \ke \r)^{-4(n+\alpha+1)}\xre^2 \l( \int_{\xre}^{x_{\nu}} dx_1 \frac{\Gk(x_\nu,x_1)}{a^2(x_1)}\r)^2 \times \l\{ F_{uu}^1(k,\ere)+F_{uu}^2(k,\ere)\r\}
\end{align}
where we defined
\begin{align}
    F_{uu}^1(k,\ere)=  \int_{u_{\rm min}}^1 \f {du} u \int_{-1}^1 d\mu f(\mu,\gamma) u^{3-2(n+\alpha)}\mJ^2_{\alpha+1/2}(u\xre)\mJ^2_{\alpha+1/2}(\xre)\label{eq:fuu1_s} \\
    F^2_{uu}(k,\ere)= \int_1^{u_{\rm max}}\frac{du}{u^4}\int_{-1}^1d\mu f(\mu,\gamma)u^{6-4(n+\alpha)}\mJ^4_{\alpha+1/2}(u\xre)\label{eq:fuu2_s}
\end{align}
 \begin{figure*}
\includegraphics[width=0.45\linewidth]{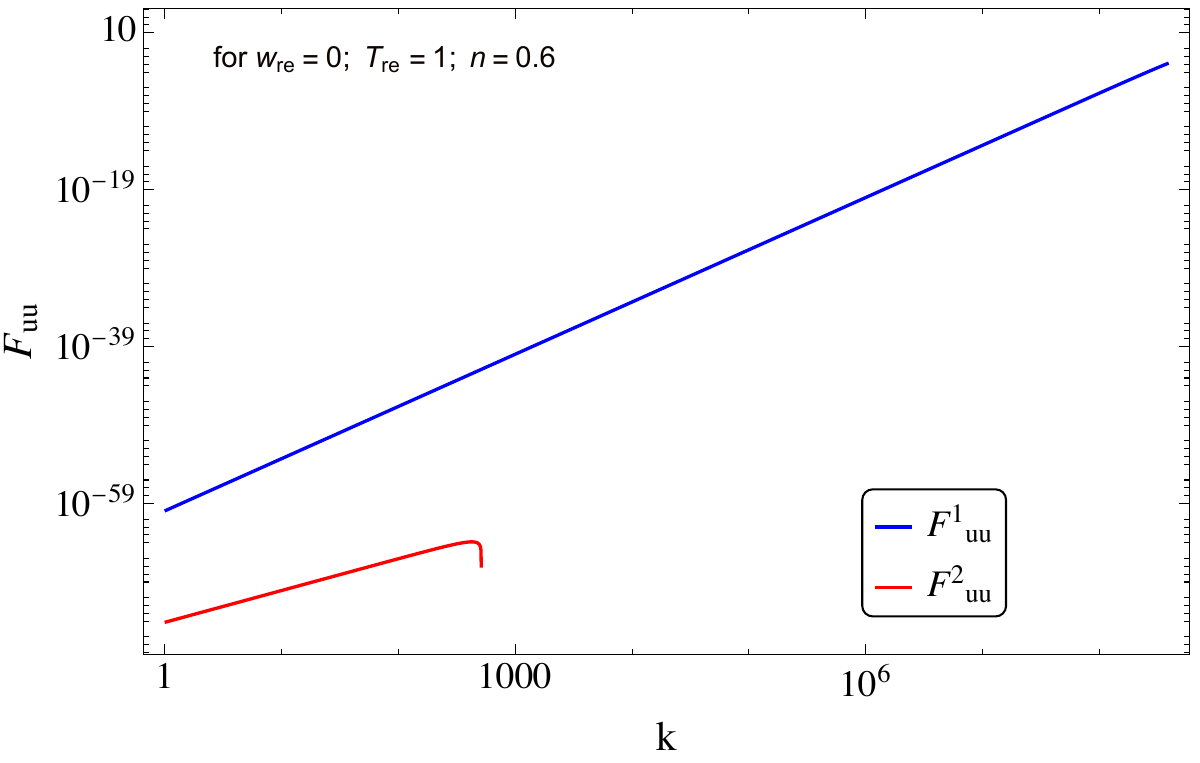}
\includegraphics[width=0.45\linewidth]{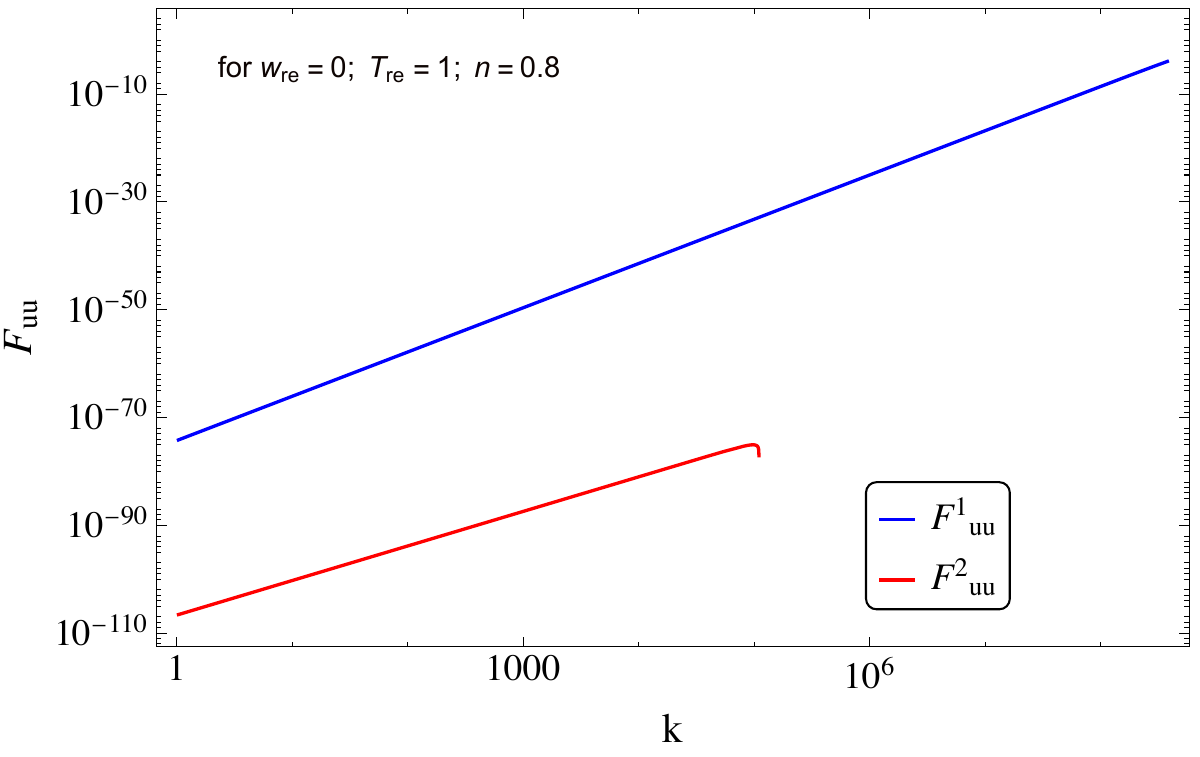}
\caption{In these figures we have plotted $F_{uu}^1$ and $F_{uu}^2$ as a function of $k$ for two specific sets of parameters and we have seen that for all cases $F_{uu}^1$ is dominated over the $F_{uu}^2$.}\label{fig:fuu}
\end{figure*}

Now, if we perform the \( \mu \)-integral in the respective limits \( u < 1 \) and \( u > 1 \), we obtain the following expressions for each regime
\begin{align}
   F_{uu}^1(k,\ere)=\frac{8}{3}  \int_{u_{\rm min}}^1 \f {du} u \int_{-1}^1 u^{3-2(n+\alpha)}\mJ^2_{\alpha+1/2}(u\xre)\mJ^2_{\alpha+1/2}(\xre)\\
    F^2_{uu}(k,\ere)=\frac{16}{15} \int_1^{u_{\rm max}}\frac{du}{u^4} u^{6-4(n+\alpha)}\mJ^4_{\alpha+1/2}(u\xre) 
\end{align}
where we used 
\begin{align}
    \lim_{u<1}\l\{ \int_{-1}^1 d\mu f(\mu,\gamma) \r\}=\int_{-1}^1 d\mu (1+\mu^2)(1+\gamma^2) \simeq \int_{-1}^1 d\mu (1+\mu^2)\simeq \frac{8}{3}\\
    \lim_{u<1}\l\{ \int_{-1}^1 d\mu f(\mu,\gamma) \r\}=\int_{-1}^1 d\mu (1+\mu^2)(1+\gamma^2)\simeq \int_{-1}^1 d\mu (1+\mu^2)^2\simeq\frac{ 16}{15}
\end{align}
\begin{align}
    \gamma=\widehat{\bf k-q} \cdot \hat{\bf k}= \frac{\vk -\vq}{|\vk-\vq|}\cdot \hat{ \vk}=\frac{\vk\cdot \hat{\vk}-\vq\cdot \hat{vk}}{|\vk-\vq|}=\frac{k-q\mu}{|\vk-\vq|}=\frac{(1-\mu u)}{|1-u|}
\end{align}
As we have seen in Fig.~\ref{fig:fuu}, the second integral (Eq.~\ref{eq:fuu2}) is subdominant compared to the first one in the range \( k < \kre \). Therefore, we can neglect the second contribution in Eq.~\ref{eq:ptra_i} and focus solely on the first integral, i.e., \( F_{uu}^1(k, \ere) \) defined in Eq.~\ref{eq:fuu1}. Although this integral has been solved numerically to study its spectral behavior, we now aim to derive an analytical estimate. Since we are interested in modes that remain outside the horizon at the end of reheating, i.e., \( k < \kre \), we can take the limit \( \xre = k \ere < 1 \) and simplify the integral further as

 \begin{figure*}
\includegraphics[width=0.5\linewidth]{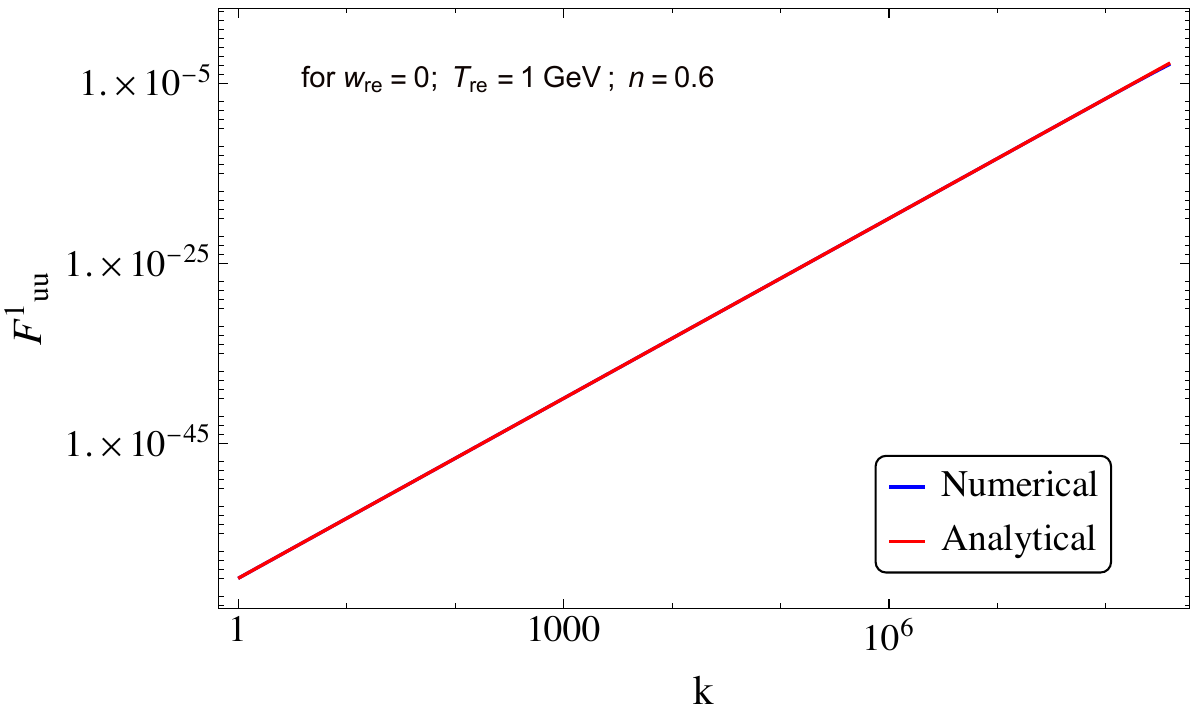}
\caption{This figure shows  
 how the numerical and analytical results are consistent for a specific set of parameters.}\label{fig:fuu_fuus}
\end{figure*}

\paragraph{\underline{Computing $F_{uu}^1(k,\ere)$ Integral:}}

\begin{align}
    F^1_{uu}(k,\ere) &=\frac{8}{3}\int_{u_{\rm min}}^1 du u^{2-2(n+\alpha)}\mJ_{\alpha+1/2}^2(u \xre) \mJ^2_{\alpha+3/2}(\xre)\\
   & =\frac{8}{3}\frac{2^{-4(\alpha+1/2)}}{\Gamma^4(\alpha+3/2)}\xre^{4(\alpha+1/2)}\int_{u_{\rm min}}^1 du u^{2(\alpha+1/2)}u^{2-2(n+\alpha)}\\
 & \simeq  \frac{8}{3}\frac{2^{-4(\alpha+1/2)}}{\Gamma^4(\alpha+3/2)}\xre^{4(\alpha+1/2)} \f 1 {4-2n} \l\{ 1 -\l( \f {k_0} {k} \r)^{4-2n}\r\}
\end{align}
n Fig.~\ref{fig:fuu_fuus}, we have plotted the numerical values along with the analytical estimation of the integral, and we have found that both are in good agreement with each other

\subsection{\underline{Computing the tensor power spectrum for $k>\kre$:}}

We now turn our attention to scenarios where all relevant modes are deep inside the horizon before the end of reheating, i.e., \( k > \kre \), which corresponds to \( \xre > 1 \). In this regime, the tensor power spectrum sourced by the magnetic field for modes \( k > \kre \) can be expressed as
\begin{align}
   \mPts(k>\kre,\eta_\nu)= \frac{k^4|\mB|^4}{32\pi^2\Mp^4}\l( \f k \ke \r)^{-4(n+\alpha+1)}\xre^2 &\l( \int_{\xre}^{x_{\nu}} dx_1 \frac{\Gk(x_\nu,x_1)}{a^2(x_1)}\r)^2\nn\\
    &\times \l\{ 
     \int_{u_{\rm min}}^1 \f {du} u \int_{-1}^1 d\mu f(\mu,\gamma) u^{3-2(n+\alpha)}\mJ^2_{\alpha+1/2}(u\xre)\mJ^2_{\alpha+1/2}(\xre) \r.\nn\\
     &\l. + \int_1^{u_{\rm max}}\frac{du}{u^4}\int_{-1}^1d\mu f(\mu,\gamma)u^{6-4(n+\alpha)}\mJ^4_{\alpha+1/2}(u\xre) \r\} \label{eq:ptra_s}
\end{align}
\begin{align}
     \mPts(k>\kre,\eta_\nu)=\frac{k^4|\mB|^4}{32\pi^2\Mp^4}\l( \f k \ke \r)^{-4(n+\alpha+1)}\xre^2 \l( \int_{\xre}^{x_{\nu}} dx_1 \frac{\Gk(x_\nu,x_1)}{a^2(x_1)}\r)^2
    \times \l\{ F_{uu}^3(k>\kre,\ere)+F_{uu}^4(k>\kre,\ere) \r\}
\end{align}
where we defined 
\begin{align}
   F_{uu}^3(k>\kre,\ere) &= \int_{u_{\rm min}}^1 \f {du} u \int_{-1}^1 d\mu f(\mu,\gamma) u^{3-2(n+\alpha)}\mJ^2_{\alpha+1/2}(u\xre)\mJ^2_{\alpha+1/2}(\xre>1) \\
   F_{uu}^4(k>\kre,\ere) &= \int_1^{u_{\rm max}}\frac{du}{u^4}\int_{-1}^1d\mu f(\mu,\gamma)u^{6-4(n+\alpha)}\mJ^4_{\alpha+1/2}(u\xre)
\end{align}
Similarly, we first perform the \( \mu \) integral and rewrite the above integral as
\begin{align}
    F_{uu}^3(k>\kre,\ere) &=\frac{8}{3} \int_{u_{\rm min}}^1 \f {du} u  u^{3-2(n+\alpha)}\mJ^2_{\alpha+1/2}(u\xre)\mJ^2_{\alpha+1/2}(\xre>1)\\
    &=\frac{8}{3}\int_{u_{\rm min}}^1 du u^{2-2(n+\alpha)} \mJ^2_{\alpha+1/2}(u\xre)\mJ^2_{\alpha+1/2}(\xre>1)\label{eq:fuu3}
\end{align}
\begin{align}
    F_{uu}^4(k>\kre,\ere) &=\frac{16}{15} \int_1^{u_{\rm max}}\frac{du}{u^4} u^{6-4(n+\alpha)}\mJ^4_{\alpha+1/2}(u\xre>>1) \\
     &=\frac{16}{15} \int_1^{u_{\rm max}}du u^{2-4(n+\alpha)}\mJ^4_{\alpha+1/2}(u\xre>>1)
\end{align}
 \begin{figure*}
\includegraphics[width=0.45\linewidth]{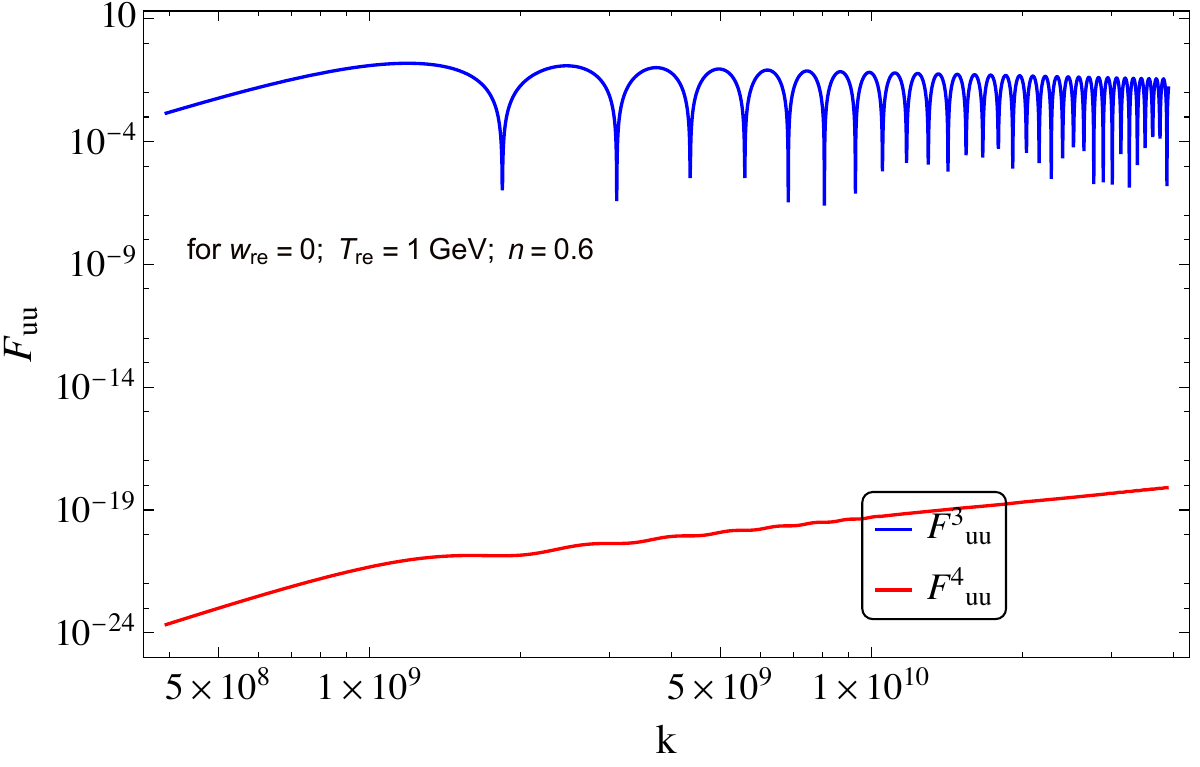}
\includegraphics[width=0.45\linewidth]{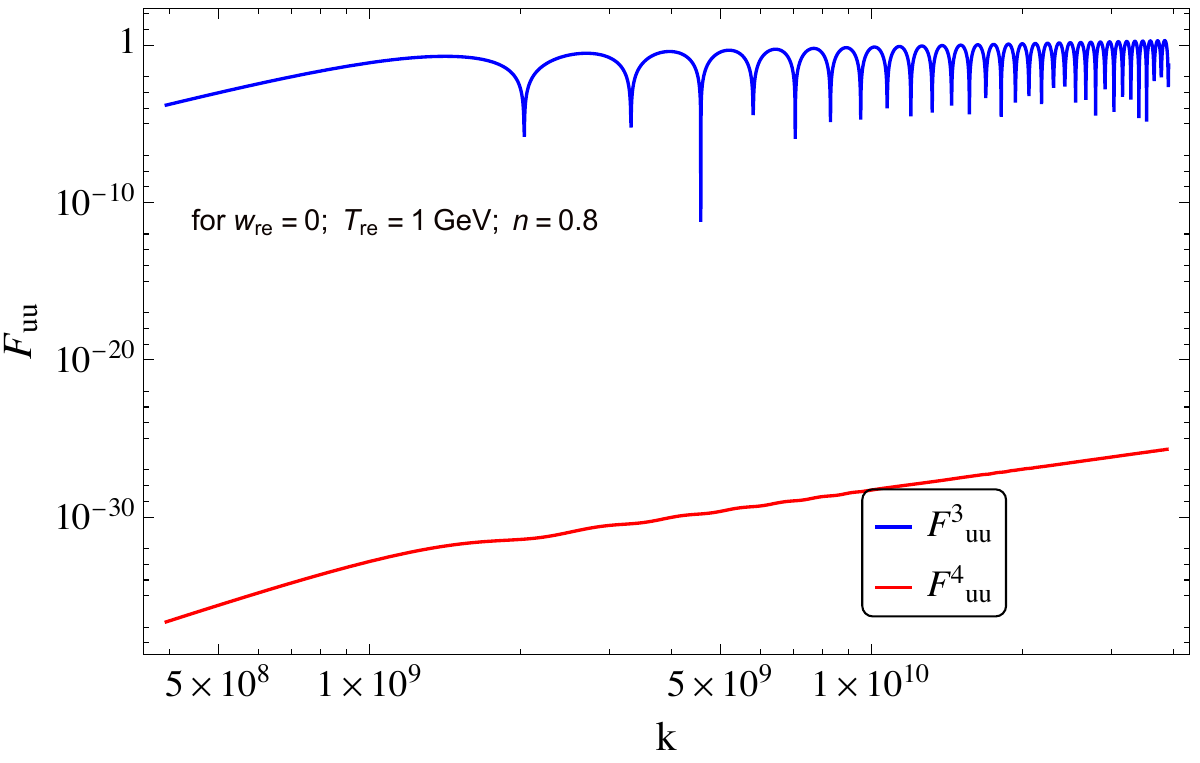}
\caption{In this figure, we have shown how the numerical and analytical estimate is consistence with each other for a specific set of parameters.}\label{fig:fu21}
\end{figure*}
From Fig.~(\ref{fig:fu21}), it is evident that for \( k > \kre \), the dominant contribution arises entirely from the first integral, i.e., \( F_{uu}^3 \) defined in Eq.~(\ref{eq:fuu3}). Hence, we neglect the second integral, \( F_{uu}^4 \), as its contribution is subdominant. To gain an analytical understanding of the GW spectral behavior, we now decompose the above integral into two parts as
\begin{align}
    F_{uu}^3(k>\kre,\ere)=\frac{8}{3}\l\{  \int_{u_{\rm min}}^{\kre/k} du u^{2-2(n+\alpha)} \mJ^2_{\alpha+1/2}(u\xre)\mJ^2_{\alpha+1/2}(\xre>1)\r.\nn\\
    \l.+ \int_{\kre/k}^1 du u^{2-2(n+\alpha)} \mJ^2_{\alpha+1/2}(u\xre)\mJ^2_{\alpha+1/2}(\xre>1) \r\}
\end{align}
\begin{align}
    F_{uu}^3(k>\kre,\ere)= F^3_{uu,1}(k>\kre,\ere)+F^3_{uu,2}(k>\kre,\ere)
\end{align}
where we defined
\begin{align}
     F^3_{uu,1}(k>\kre,\ere)= \frac{8}{3} \int_{u_{\rm min}}^{\kre/k} du u^{2-2(n+\alpha)} \mJ^2_{\alpha+1/2}(u\xre)\mJ^2_{\alpha+1/2}(\xre>1)\\
     F^3_{uu,2}(k>\kre,\ere)=\f 8 3 \int_{\kre/k}^1 du u^{2-2(n+\alpha)} \mJ^2_{\alpha+1/2}(u\xre)\mJ^2_{\alpha+1/2}(\xre>1)
\end{align}
 \begin{figure*}
\includegraphics[width=0.45\linewidth]{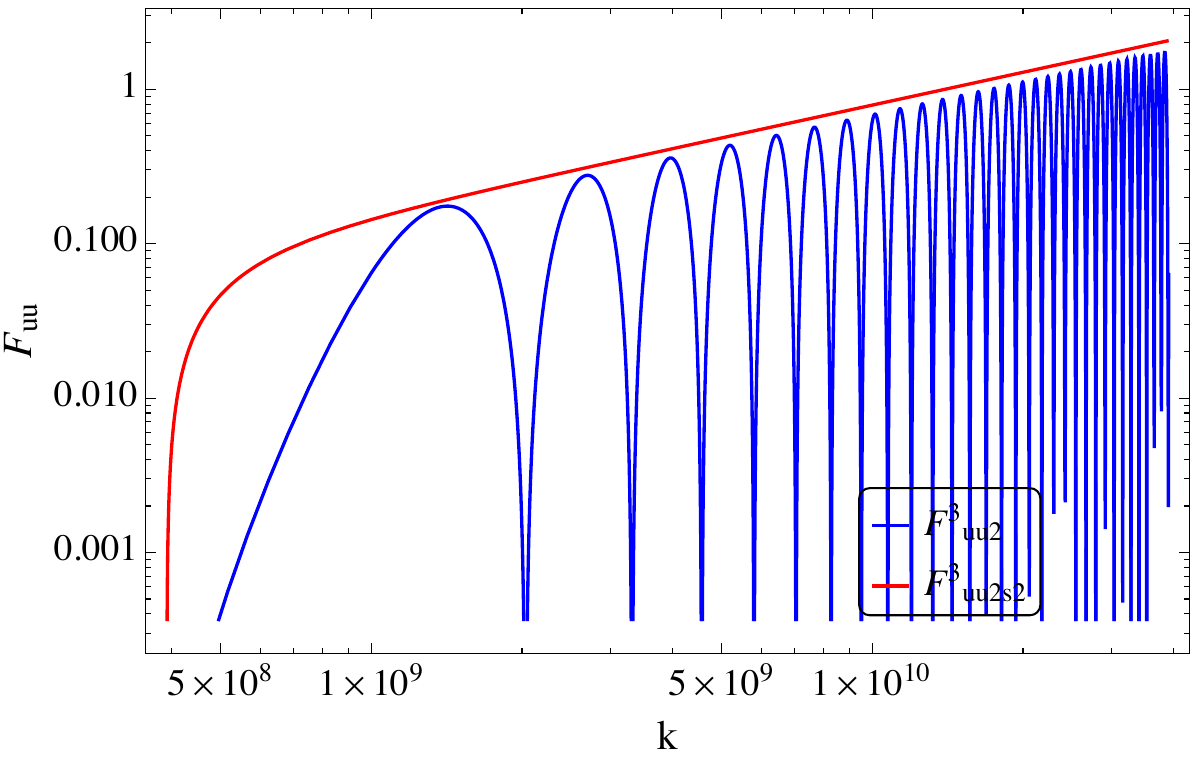}
\includegraphics[width=0.45\linewidth]{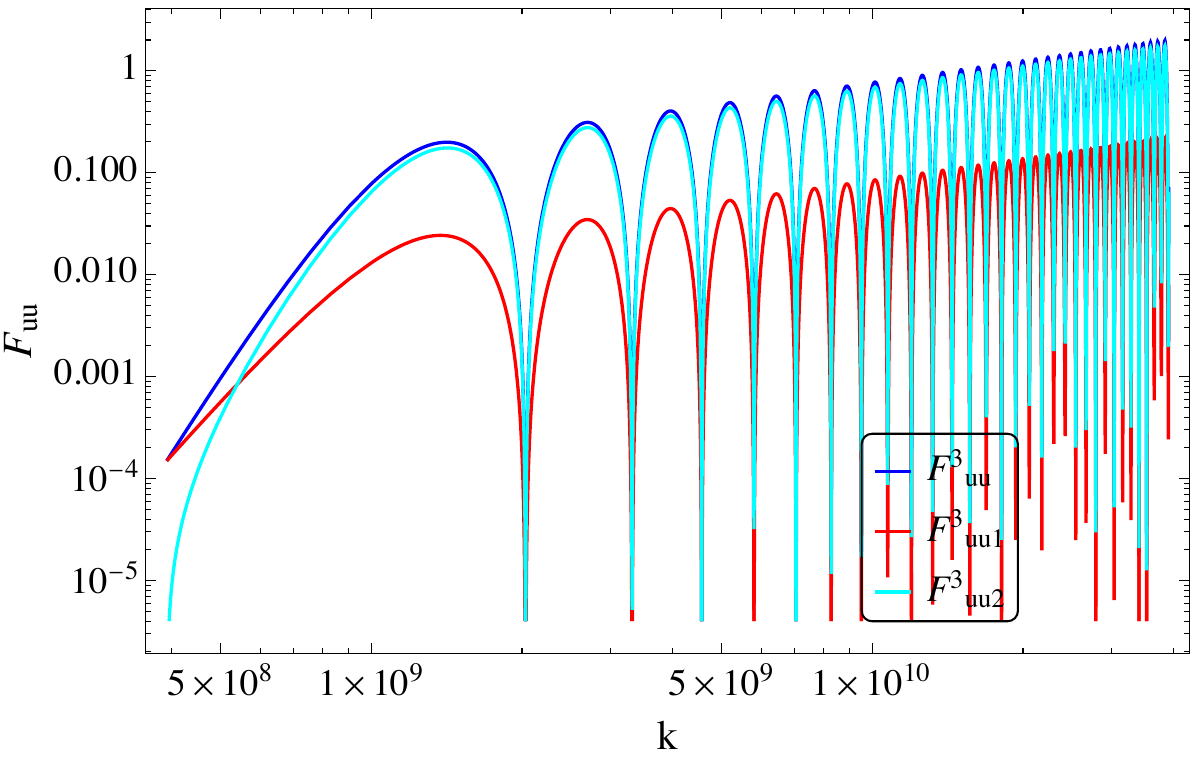}
\caption{In this figure, we have shown how the numerical and analytical estimates are consistent for a specific set of parameters.}\label{fig:fuu_fuu2}
\end{figure*}
As shown in Fig.~\ref{fig:fuu_fuu2}, we find that the last integral provides the best fit to the full numerical result. Therefore, we retain only the last integral for further analytical estimations. Since we are focusing on the modes that remain inside the horizon at the end of reheating, i.e., \( k > \kre \), and the integration range extends from \( \kre/k \) to 1, the above expression can be further simplified as
\begin{align}
   F^3_{uu,2}(k>\kre,\ere)=\f 8 3 \f 4 {\pi^2} \f 1 {\xre^2} \int_{\kre/k}^1 du u^{1-2(n+\alpha)}\\
   =\f 8 3 \f 4 {\pi^2} \f 1 {\xre^2} \frac{1}{2-2(n+\alpha)}\l\{ 1- \l( \frac{\kre}{k}\r)^{2-2(n+\alpha)}\r\}
\end{align}
However, as we have seen, to ensure consistency with the full numerical amplitude, we need to include an overall numerical prefactor. We find that this prefactor is approximately \( 0.2 \). Therefore, the final analytical expression for the integral becomes
\begin{align}
   F^3_{uu}(k>\kre)\simeq \frac{64}{30\pi^2}\frac{1}{2-2(n+\alpha)}\frac{1}{\xre^2}\l\{ 1- \l( \frac{\kre}{k}\r)^{2-2(n+\alpha)}\r\}
\end{align}

\bibliographystyle{apsrev4-1}
\bibliography{references}
\end{document}